 \documentclass[10pt,preprint]{aastex}  
 \usepackage{natbib}
\def\ang{\AA}

\def\gapprox{\lower.4ex\hbox{$\;\buildrel >\over{\scriptstyle\sim}\;$}}
\def\lapprox{\lower.4ex\hbox{$\;\buildrel <\over{\scriptstyle\sim}\;$}}

\shortauthors{ASCHWANDEN ET AL. 2016}
\shorttitle{Global Energetics of Solar Flares. III.}

\begin{document}

\title{         Global Energetics of Solar Flares: 
		III. Nonthermal Energies }

\author{        Markus J. Aschwanden$^1$}

\affil{		$^1)$ Lockheed Martin, 
		Solar and Astrophysics Laboratory, 
                Org. A021S, Bldg.~252, 3251 Hanover St.,
                Palo Alto, CA 94304, USA;
                e-mail: aschwanden@lmsal.com}

\author{	Gordon Holman$^2$}

\affil{		$^2)$ Code 671, NASA Goddard Space Flight Center,
		Greenbelt, MD 20771, USA;
		e-mail: gordon.d.holman@nasa.gov }

\author{        Aidan O'Flannagain$^3$}

\affil{		$^3)$ Astrophysics Research Group,
		School of Physics, Trinity College Dublin,
		Dublin 2, Ireland;
 		e-mail: aidanoflann@gmail.com}

\author{	Amir Caspi$^4$}

\affil{		$^4)$ Planetary Science Directorate,
		Southwest Research Institute,
		Boulder, CO 80302, USA;
		e-mail: amir.caspi@swri.org }

\author{	James M. McTiernan$^5$}

\affil{		$^5)$ Space Sciences Laboratory,
		University of California, 
		Berkeley, CA 94720, USA;
		e-mail: jimm@ssl.berkeley.edu } 

\and

\author{	Eduard P. Kontar$^6$}

\affil{		$^6)$ School of Physics and Astronomy, University of Glasgow,
    		G12 8QQ, Glasgow, Scotland, UK;
		e-mail: eduard.kontar@astro.gla.ac.uk }

\begin{abstract}
This study entails the third part of a global flare energetics project,
in which {\sl Ramaty High-Energy Solar Spectroscopic Imager (RHESSI)} data
of 191 M and X-class flare events from the first 3.5 yrs of the {\sl Solar 
Dynamics Observatory (SDO)} mission are analyzed. We fit a thermal and a
nonthermal component to RHESSI spectra, yielding the temperature
of the differential emission measure (DEM) tail, the nonthermal power law 
slope and flux, and the thermal/nonthermal cross-over energy $e_{\mathrm{co}}$. 
From these parameters we calculate the total nonthermal energy $E_{\mathrm{nt}}$ 
in electrons with two different methods: (i) using the observed cross-over 
energy $e_{\mathrm{co}}$ as low-energy cutoff, and (ii) using the low-energy 
cutoff $e_{\mathrm{wt}}$ predicted by the warm thick-target bremsstrahlung model
of Kontar et al. {\bf Based on a mean temperature of $T_e=8.6$ MK in active regions
we find low-energy cutoff energies of $e_{\mathrm{wt}} =6.2\pm 1.6$ keV for 
the warm-target model, which is significantly lower than the cross-over
energies $e_{\mathrm{co}}=21 \pm 6$ keV. Comparing with the statistics of 
magnetically dissipated energies $E_{\mathrm{mag}}$ and thermal energies 
$E_{\mathrm{th}}$ from the two previous studies, we find the following mean
(logarithmic) energy ratios with the warm-target model:
$E_{\mathrm{nt}} = 0.41 \ E_{\mathrm{mag}}$, 
$E_{\mathrm{th}} = 0.08 \ E_{\mathrm{mag}}$, and 
$E_{\mathrm{th}} = 0.15 \ E_{\mathrm{nt}}$.  
The total dissipated magnetic energy exceeds the thermal energy in 95\%
and the nonthermal energy in 71\% of the flare events, which confirms
that magnetic reconnection processes are sufficient to explain 
flare energies. The nonthermal energy exceeds the thermal energy in 85\% 
of the events, which largely confirms the warm thick-target model.}
\end{abstract}

\section{		    INTRODUCTION			}

We undertake a systematic survey of the global energetics of solar flares and
coronal mass ejections (CME) observed during the SDO era, which includes
all M and X-class flares during the first 3.5 years of the SDO mission,
covering some 400 flare events.  This project embodies the most comprehensive
survey about various forms of energies that can be detected during flares, such
as the dissipated magnetic energy, the thermal energy, the nonthermal
energy, the radiative and conductive energy, and the kinetic energy
of associated CMEs. Two studies have been completed previously,
containing statistics on magnetic energies (Aschwanden Xu, and Jing 2014;
Paper I), and thermal energies (Aschwanden et al.~2015; Paper II).
In this study we focus on the third
part of this ``global flare energetics project", which entails the statistics
of nonthermal energies in hard X ray-producing electrons 
that are observed in hard X-rays and gamma-rays, using data from the
Ramaty High-Energy Solar Spectroscopic Imager (RHESSI) spacecraft
(Lin et al.~2002).

The quantitative measurement of nonthermal energies in solar flares 
allows us some tests of fundamental nature. One concept or working hypothesis 
is that all primary energy input in solar flares is provided by dissipation 
of free magnetic energy, for instance by a magnetic reconnection process,
which supplies energy for secondary processes, such as for acceleration 
of charged particles and heating of flare plasma. 
The accelerated (nonthermal) particles either escape
from the flare site into interplanetary space, or more likely precipitate
down to the chromosphere where they subsequently become thermalized and 
radiate in hard X-rays and gamma rays, according to the thick-target
bremsstrahlung model (Brown 1971). In this picture we expect that the 
total nonthermal energy $E_{\mathrm{nt}}$ (in electrons and ions) 
produced in flares should not exceed the dissipated magnetic (free) 
energy $E_{\mathrm{mag}}$, but on the other hand should yield an 
upper limit on the thermal energy $E_{\mathrm{th}}$ 
inferred from the soft-X-ray and EUV-emitting plasma. 
Alternative mechanisms to the thick-target model envision 
thermal conduction fronts (e.g., Brown et al.~1979) or
direct heating processes (e.g., Duijveman et al.~1981).
In the previous
two papers we proved the inequality $E_{\mathrm{mag}} > E_{\mathrm{th}}$, 
for which we found an energy conversion ratio of 
$E_{\mathrm{th}}/E_{\mathrm{mag}} \approx 0.02-0.40$
(Paper II), which is about an order of magnitude higher than estimated
in a previous statistical study (Emslie et al.~2012), where an 
{\sl ad hoc} value (30\%) of the ratio of the free magnetic energy to the
potential field energy was estimated. In this Paper III we investigate the expected 
inequalities $E_{\mathrm{mag}} > E_{\mathrm{nt}} > E_{\mathrm{th}}$.
If these two inequalities are not fulfilled, it could be attributed 
to insufficient accuracy of the energy measurements, or alternatively may question 
the correctness of the associated low-energy cutoff model, the applied 
magnetic reconnection models, or the efficiency of the electron
thick-target bremsstrahlung model. 
Such an outcome would have important consequences in our understanding 
of solar flare models and the related predictability of the most extreme 
space weather events.

The measurement of nonthermal energies in solar flares requires
a spectral fit of the hard X-ray spectrum in the energy range of
$\varepsilon \approx 10-30$ keV (Aschwanden 2007), 
from spectral data as they are available from
the HXRBS/SMM, BATSE/CGRO, or RHESSI instrument. Since the total
nonthermal energy contained in a flare requires integrations over
the temporal and spectral range, the largest uncertainty of this
quantity comes from the assumed low-energy cutoff,  
because it cannot be directly measured due to the strong thermal
component that often dominates the spectrum at $\varepsilon \lapprox 20$ keV during
solar flares (for a review see Holman et al.~2011).
In a few cases, low-energy cutoffs of the nonthermal spectrum
could be determined by regularized inversion methods at $e_c=20-40$ keV 
(Kasparova et al.~2005), $e_c \approx 20$ keV (Kontar and Brown 2006),
and $e_c=13-19$ keV (Kontar, Dickson, and Kasparova 2008).
For the 2002 July 23 flare, Holman et al.~(2003) deduced upper limits to 
low-energy cutoffs by determining the highest values consistent with 
acceptable spectral fits.  Sui et al.~(2007) deduced the low-energy cutoff 
in a flare from the combination of spectral fits and the time evolution of 
the X-ray emission in multiple energy bands. Sui et al.~(2007) deduced 
low-energy cutoffs for several flares with relatively weak thermal 
components (``early impulsive flares'') from spectral fits, with 
values ranging from $15-50$ keV. In the late peak of a multi-peaked flare, 
Warmuth et al.~(2009) inferred low-energy cutoff values exceeding 100 keV,
but this unusually high value could possibly be explained also by 
high-energy electrons that accumulate by trapping after the flare peak
(Aschwanden et al.~1997). Using a novel method of {\bf differentiating}
nonthermal
electrons by their time-of-flight delay from thermal electrons by their
thermal conduction time delay, a thermal-nonthermal crossover energy of
$e_c=18.0 \pm 3.4$ keV (or a range of $e_c = 10-28$ keV) was established
for the majority (68\%) of 65 analyzed flare events (Aschwanden 2007). 

Statistical measurements of nonthermal flare energies 
have been calculated from HXRBS/SMM data (Crosby et al.~1993),
or from RHESSI data (Hannah et al.~2008; Christe et al.~2008; 
Emslie et al.~2012). 
The low-energy cutoff was taken into account by assuming a
fixed energy cutoff of $e_c=25$ keV (Crosby et al.~1993), 
a fixed spectral slope of $\gamma=-1.5$ below the thermal-nonthermal
cross-over energy $e_{\mathrm{co}}$ (Hannah et al.~2008), 
or by adopting the largest energy $e_c$ that still produces a 
goodness-of-fit with $\chi^2 \approx 1$ for the nonthermal
power law fit (Emslie et al.~2012). Low-energy cutoffs for
microflares were estimated in the range of $e_c \approx 9-16$ keV,
with a median of 12 keV (Hannah et al.~2008), using a numerical
integration code of Holman (2003).
The statistical study of
Emslie et al.~(2012) provides a comparison between nonthermal energies
$E_{\mathrm{nt}}$, thermal energies $E_{\mathrm{th}}$, and dissipated magnetic 
energies $E_{\mathrm{mag}}$, yielding mean (logarithmic) ratios of 
$E_{\mathrm{th}} \approx 0.005\ E_{\mathrm{mag}}$ and $E_{\mathrm{nt}} 
\approx 0.03\ E_{\mathrm{mag}}$.  
These results conform to the expected inequalities, but the magnetic 
energies $E_{\mathrm{mag}}$ were actually not measured in the study of 
Emslie et al.~(2012), 
and most likely were overestimated by an order of magnitude (Paper I).
The dissipated magnetic energies $E_{\mathrm{mag}}$ were for the first time 
quantitatively measured in Paper I, by automated tracing of coronal
flare loops from AIA/SDO images and by forward-fitting of a nonlinear
force-free magnetic field (NLFFF) model based on the vertical current
approximation (Aschwanden 2013, 2016).

The content of this paper consists of a theoretical model
to estimate the low-energy cutoff and the nonthermal energy (Section 2),
a description of the data analysis method (Section 3), the results
of the data analysis of 191 M and X-class flare events observed 
with RHESSI (Section 4), a discussion of the results (Section 5), 
and conclusions (Section 6). 

\section{		    THEORY 				}

\subsection{	Nonthermal Energy in Electrons			}

The nonthermal energy in flare electrons is generally calculated
with the thick-target model (Brown 1971), which expresses the
hard X-ray photon spectrum by a convolution of the electron
injection spectrum with the Bethe-Heitler bremsstrahlung
cross-section. According to this model, the observed hard X-ray
photon spectrum $I(\varepsilon)$ observed at Earth can be approximated
by a power law function with slope $\gamma$ for the nonthermal
energies, while the spectral index generally changes at the lower
(thermal) energies. Thus, the nonthermal spectrum is defined as
(e.g., see textbook Aschwanden 2004; chapter 13),
\begin{equation}
        I(\varepsilon) = A \ \varepsilon^{-\gamma}
        \qquad ({\rm photons}\ {\rm cm}^{-2}\ {\rm s}^{-1}\ {\rm keV}^{-1}) \ ,
\end{equation}
which yields a thick-target (non-thermal) electron injection 
spectrum $f_e(e)$,
\begin{equation}
        f_e(e) = 2.68 \times 10^{33} \ b(\gamma) A
        {e}^{-(\gamma+1)}
        \qquad ({\rm electrons}\ {\rm keV}^{-1} \ {\rm s}^{-1}) \ ,
\end{equation}
which is a power law function also, but with a slope $\delta = \gamma+1$
that is steeper by one, and $b(\gamma)$ is an auxiliary function related
to the Beta function. The detailed shape of a nonthermal electron spectrum
that is affected by a low-energy cutoff is simulated in Holman (2003),
showing a gradual flattening at lower energies.
Note that we use the symbol $\varepsilon$ for photon
energies, while we use the symbol $e$ for electron energies.
The total power in nonthermal electrons above some cutoff
energy $e_c$, i.e., $P(e \ge e_c$), is
\begin{equation}
        P(e \ge e_c)
        = 4.3 \times 10^{24} \ {b(\gamma) \over (\gamma - 1)} \ A \
        (e_c)^{-(\gamma-1)} \
	\qquad ({\rm erg \ s^{-1})} \ .
\end{equation}
Thus, the three observables of the photon
flux $A$, the photon power law slope $\gamma$, and the low-energy
cutoff energy $e_c$ are required to calculate the
power during a selected flare time interval, which can be
calculated with the OSPEX package of the {\sl SolarSoftWare (SSW)} library
of the {\sl Interactive Data Language (IDL)} software (see RHESSI webpage
http://hesperia.gsfc.nasa.gov/ssw/packages/spex/doc/ospex$\_$explanation.html).

In order to calculate the total
nonthermal energy $E_{\mathrm{nt}}$ during an entire flare, we have to
integrate the power as a function of time,
\begin{equation}
        E_{\mathrm{nt}} = \int_{t_{\mathrm{start}}}^{t_{\mathrm{end}}} 
	P(e > e_c(t), t) \ dt 
	\qquad ({\rm erg}) \ .
\end{equation}
While the photon fluxes $A(t)$ and the spectral slopes $\gamma(t)$ can
readily be measured from a time series of hard X-ray photon spectra (Eq.~1),
the largest uncertainty in the determination of the nonthermal energy
is the low-energy cutoff energy $e_c(t)$ between the thermal
and nonthermal hard X-ray components, typically expected in the range of 
$\approx 10-30$ keV (see Table 3 in Aschwanden 2007). 
In the following we outline two different
theoretical models of the low-energy cutoff that are applied 
in this study.

\subsection{		Thermal-Nonthermal Cross-Over Energy 		}

The bremsstrahlung spectrum $I(\varepsilon)$ of a thermal plasma with
temperature $T$, as a function of the photon energy $\varepsilon = h\nu$,
setting the coronal electron density equal to the ion density $(n=n_i=n_e)$,
and neglecting factors of order unity (such as the Gaunt factor $g(\nu, T)$ 
in the approximation of the Bethe-Heitler bremsstrahlung cross-section),
and the ion charge number, $Z\approx 1$, is (Brown 1974; Dulk \& Dennis 1982),
\begin{equation}
        I(\varepsilon) = I_0 \int
        {\exp{(-{\varepsilon / k_B T})} \over T^{1/2}} {dEM(T) \over dT} \ dT \ ,
\end{equation}
where $I_0 \approx 8.1 \times 10^{-39}$ keV s$^{-1}$ cm$^{-2}$ keV$^{-1}$ and
$dEM(T)/dT$ specifies the {\sl differential emission measure}
$n^2 dV$ in the element of volume $dV$ corresponding to temperature range $dT$,
\begin{equation}
        \left({dEM(T) \over dT}\right) dT = n^2(T) \ dV \ .
\end{equation}
Regardless, whether we define this differential emission measure (DEM)
distribution by an isothermal or by a multi-thermal plasma (Aschwanden 2007), 
the thermal spectrum $I(\varepsilon)$ falls off similar to an exponential 
function at an energy of $\varepsilon \lapprox 20$ keV (or up to $\lapprox
40$ keV in extremal cases), while the nonthermal 
spectrum in the higher energy range of $\varepsilon \approx 20-100$ keV
can be approximated with a single (or broken) power law function (Eq.~1).

Because of the two different functional shapes, a cross-over energy 
$\varepsilon_c$ can often be defined from the change in the spectral 
slope between the thermal and the nonthermal spectral component. 
The electron energy spectrum, however, can have a substantially lower 
or higher cutoff energy (e.g., Holman 2003). 

We represent the combined spectrum with the sum of the (exponential-like)
thermal and the (power law-like) nonthermal component, i.e., 
\begin{equation}
	I(\varepsilon) =I_{\mathrm{th}}(\varepsilon)+I_{\mathrm{nt}}(\varepsilon)
	= I_0 \int {\exp{(-{\varepsilon / k_B T})} \over T^{1/2}} 
	{dEM(T) \over dT} \ dT \ + \ A \ \varepsilon^{-\gamma} \ ,
\end{equation}
where the cross-over energy $\varepsilon_{\mathrm{co}}$ can be determined in the
(best-fit) model spectrum $I(\varepsilon)$ from the energy where the logarithmic
slope is steepest, i.e., from the maximum of $\partial \log I(\varepsilon) 
/ \partial \log \varepsilon$.

\subsection{		Warm-Target Model		}

A new theoretical model has recently been developed that allows us to
calculate the low-energy cutoff energy in the thick-target model
directly, by including the ``warming'' of the cold thick-target plasma
during the electron precipitation phase, when chromospheric
heating and evaporation sets in (Kontar et al.~2015). 
Previous applications of the thick-target model generally assume
cold (chromospheric) temperatures in the electron precipitation site
(e.g., Holman et al.~2011, for a review).
The theoretical derivation of the warm-target model has been analytically
derived and tested with numerical simulations that include the effects of
collisional energy diffusion and thermalization of fast electrons
(Galloway et al.~2005; Goncharov et al.~2010; Jeffrey et al.~2014).
According to this model, the effective low-energy cutoff $e_c$ is
a function of the temperature $e_{\mathrm{th}}=k_B T_e$ of the warm-target
plasma and the power law slope $\delta = \gamma + 1$ of the 
(nonthermal) electron flux,
\begin{equation}
        e_c \approx (\xi + 2) \ k_B T_e = \delta \ k_B T_e \ .
\end{equation}
where $\xi = \gamma - 1$ is the power law slope of the 
source-integrated mean electron flux spectrum ({\bf see Eqs.~8-10
in Kontar et al.~2015), and $T_e$ is the temperature of the
warm target, which is a mixture or the cold preflare plasma and
the heated evaporating plasma.}
Thus, for the temperature range of a medium-sized to a large
X-class flare, which spans $T_e \approx 10-25$ MK, the temperature in
energy units is $E_{\mathrm{th}} = k_B T_e \approx 0.9-2.1$ keV, 
and for a range of
power law slopes of $\delta = 3-6$ (Dennis 1985; Kontar et al.~2011),
a range of $e_c \approx 3-13$ keV is predicted for the low-energy cutoffs
by this model.

Besides collisional heating of the warm chromospheric target,
electron beams and beam-driven Langmuir wave turbulence may affect
the low-energy cutoff additionally (Hannah et al.~2009). 
Alternative analytical models on the low-energy cutoff can be derived
from a collisional time-of-flight model (Appendix A), from the
Rosner-Tucker-Vaiana heating/cooling balance model (Appendix B), and 
from the runaway acceleration model (Appendix C).

\section{ 		DATA ANALYSIS METHOD				}

From the same comprehensive catalog of 399 M and X-class flares observed with
SDO during 2010-2014, used in the first two studies of our global flare
energetics project, we will analyze all events that have been simultaneously
observed in hard X-rays and gamma-rays with RHESSI. The orbit of
RHESSI has a duty cycle of $\approx 50\%$, leading to a total of
191 events that have suitable time coverage. In the following we describe
the analysis of these events, which are also listed in Table 1  
(labeled with identical identification numbers \#1$-$399 as used in 
Papers I and II). We explain the various steps performed in our analysis 
for three examples shown in Figs.~1-3. 

\subsection{Spectral Modeling of RHESSI Data with OSPEX }

For the measurement of the nonthermal energy ($E_{\mathrm{nt}}$) of electrons
during solar flares we use the OSPEX (Object Spectral Executive) software,
which is an object-oriented interface for X-ray spectral analysis of
solar data, written by Richard Schwartz and others (see RHESSI website 
{\sl http://hesperia.gsfc.nasa.gov/} for a documentation).
The OSPEX software allows the user to read RHESSI data,
to select and subtract a background, to select time intervals of interest,
to select a combination of photon flux model components,
and to fit those components to the spectrum in each selected time
interval.  During the fitting process, the response matrix is
used to convert the photon model to the model counts that are fitted 
to the observed counts. The OSPEX software deals also with changes of
attenuator states, decimation, pulse pile-up effects, albedo effects,
and provides procedures to calculate the nonthermal energy $(E_{\mathrm{nt}})$ 
(according to the thick-target model) and the thermal energy ($E_{\mathrm{th}}$) 
down to energies of $\gapprox 3$ keV. 

RHESSI complements spectral
information of the differential emission measure (DEM) distribution
at the high-temperature side ($T_e \gapprox 16$ MK) 
(Caspi 2010; Caspi and Lin 2010; Caspi et al.~2014),
while AIA/SDO provides DEM information at the low-temperature side
($T_e \lapprox 16$ MK), as we determined in Paper II. For spectral modeling 
we are using the two-component model {\sl vth+thick2$\_$vnorm}, which includes a 
thermal component at low energies and a (broken) power law function at higher 
(nonthermal) energies. In our spectral fits we are only interested in
the transition from the thermal to the nonthermal spectrum, which can be
expressed by an exponential-like plus a single power law function (Eq.~7), 
and thus we use only the lower power law part of the two-component model 
{\sl vth+thick2$\_$vnorm}, while the spectral slope in the upper part was set to 
a constant ($\delta_2=4$). In addition we use
{\sl calc$\_$nontherm$\_$electron$\_$energy$\_$flux} of the OSPEX 
package to calculate the nonthermal energy flux in the thick-target model.

\underbar{RHESSI Spectral Fitting Range Selection:}
In order to obtain a self-consistent measure of the nonthermal energy, which
varies considerably during the duration of a flare or among different flares,
we have to choose a spectral fitting range that covers a sufficient
part of both the thermal and nonthermal components. {\bf We choose
the maximum energy range $[\varepsilon_1,\varepsilon_2]$, bound by 
$\varepsilon_1 = 6 ... 10$ keV and $\varepsilon_2 = 20 ... 50$ keV,
in which an acceptable (reduced) $\chi^2$-value ($\chi < 2.0$) is obtained 
for the spectral fit. 
The upper bound of the fitting range is mostly constrained by the photon 
count statistics, which is often too noisy for energies at $\varepsilon_2 
\gapprox 30$ keV during small flares (M-class here), given the
time steps of $\Delta t=20$ s chosen throughout.} The fitted energy ranges 
cover also the range of cross-over energies (10-28 keV) found in
multi-thermal fitting of energy-dependent time delays (Aschwanden 2007). 

As a general criticism, we have to be aware that the nonthermal
spectral component could in addition also be confused with a multi-thermal 
component in the fitted spectral range of $\varepsilon \approx 10-30$ keV 
(Aschwanden 2007), or with non-uniform ionization effects (Su, Holman, and 
Dennis 2011), or with return-current losses (Holman 2012).

\medskip
\underbar{RHESSI Detector Selection:}
We used the standard option of OSPEX, where a spectral fit is calculated 
from the combined counts of a selectable set of RHESSI subcollimaters.
RHESSI has 9 (subcollimator) detectors that had originially near-identical
sensitivities, but progressively deviate from each other as a result of 
steady degradation over time due to radiation damage from charged particles.
Heating up the germanium restores the lost sensitivity and resolution,
and thus five annealing procedures have been applied to RHESSI so far
(second anneal at 2010 Mar 16 - May 1; third at 2012 Jan 17 - Febr 22; 
forth at 2014 Jun 26 - Aug 13; and fifth at 2016 Feb 23 - Apr 23).
No science data are collected during the annealing periods. Based on
the performance of the individual detector sensitivities, it is
general practice to exclude the detectors 2 and 7 in spectral fits.
Furthermore, detectors 4 and 5 are considered as unreliable {\bf after
January 2012} (Richard Schwartz; private communication). Therefore, we select
the set of detectors [1, 3, 4, 5, 6, 8, 9] in spectral fits up to 
the third anneal in January 2012 (events \# 1-126 in Table 1), and the set of 
[1, 3, 6, 8, 9] after February 2012 (events \# 154-395 in Table 1). 
Omitting detectors 4 and 5 in the latter set of 71 events yields
a total nonthermal energy that is by a factor of $q_{\mathrm{det}}=
E_{\mathrm{nt}}[1, 3, 6, 8, 9]/E_{\mathrm{nt}}[1, 3, 4, 5, 6, 8, 9]=1.3 \pm 0.5$
higher. 

\medskip
\underbar{GOES Time Range and RHESSI Time Resolution:}
We download the GOES 1-8 \ang\ light curves $F_{\mathrm{GOES}}(t)$ and
calculate the time derivative as a proxy for the hard X-ray
time profile $F_{\mathrm{HXR}} \approx dF_{\mathrm{GOES}}/dt$, as shown in Figs.~1a ,2a, and 3a.
The start time $t_{\mathrm{start}}$, peak time $t_{\mathrm{peak}}$,
and end time $t_{\mathrm{end}}$ are defined from the NOAA/GOES catalog.
We compute consecutive spectra in time steps of $\Delta t=20$ s.
Note that RHESSI is a spinning spacecraft with a period of 4 s,
which does not cause any modulation effects for 20 s time integrations. 

\medskip
\underbar{RHESSI Quick-Look Data:}
In a next step we inspect the RHESSI quick-look time profiles 
(Figs.~1b, 2b, 3b), which show photon counts in
{\bf 5 different energy channels in the range of 6-300 keV}. Based on
these RHESSI time profiles we select time intervals for background
subtraction. {\bf Generally we select a time interval at flare start 
as the background interval (in 90\%), and subtract this preflare spectrum
for the entire flare time interval. Only in a few cases (10\%)
where the preflare flux is higher than the postflare flux, we choose
a time interval at flare end for background subtraction.}
The RHESSI quick-look data show changes in the attenuator state
(e.g., Figs.2b, 3b),
which are automatically handled in most time intervals with the OSPEX
software, unless there is a change in the attenuator state during a
selected time interval itself, in which case this time interval is removed
from the spectral analysis. The quick-look data show occasionally data gaps
that are caused when RHESSI enters
spacecraft night in its near-Earth orbit. If the data gap does not 
occur during the flare peak of hard X-ray emission, we still include 
the event in the analysis, as long as the time interval of dominant
nonthermal HXR emission is covered (such as event \#219 in Fig.~2b).

\medskip
\underbar{OSPEX Spectral Fitting:}
For spectral fitting we perform first a semi-calibration
and store the detector response matrix (DRM), and then
run a spectral fit with the fit function {\sl vth+thick2$\_$vnorm}
using the OSPEX software, optimizing the following model fit
parameters (for each time interval $t$):

\begin{tabular}{ll}
$EM(t)$     & = Emission measure in units of $10^{49}$ cm$^{-3}$ \\
$T_e(t)$    & = plasma temperature in units of keV (1 keV=11.6 MK) \\
$A(t)$      & = photon flux at $\varepsilon = 50$ keV \\
$\delta(t)$ & = negative power law index of electron spectrum \\
$e_c(t)$    & = low-energy cutoff \\
\end{tabular}

Examples of spectral fits are shown in 
Figs.~1c, 2c, and 3c, fitted at the time of the peak power
$P_{\mathrm{co}}(t)$ (indicated with red vertical lines in Figs.~1, 2, and 3). 
The best-fit spectrum yields a cross-over energy $e_{\mathrm{co}}$ between
the thermal and nonthermal spectral component. Alternatively,
the warm-target model of Kontar et al.~(2015) yields a low-energy
cutoff value $e_{\mathrm{wt}}$. The fitted energy ranges 
are listed in Table 1 and are indicated with dotted vertical lines 
in Figs.~1g, 2g, and 3g. The goodness-of-fit is quantified with the 
$\chi^2$-value criterion. 
In case of bad fits of the $\chi^2$-values ($\chi 
> 2$) {\bf we changed either the fitted energy range (in 13\%), 
the selected interval for background subtraction (10\%),
or the fitted time range (5\%).}

\section{		RESULTS 	 			}

The numerical values of the main results of the low energy cutoffs
$e_c$ (which we label as $e_{\mathrm{co}}$ in the cross-over method, 
and as $e_{\mathrm{wt}}$ in the warm-target method), and the nonthermal 
energy $E_{\mathrm{nt}}$ for the analyzed 191 events are listed in 
Table 1, while scatter plots and distributions are shown in Figs.~4-8.

\subsection{	Time Evolution of Flares			}

Three examples of analyzed flare events are shown in Figs.~1, 2, and 3,
including one of the smallest events (Fig.~1: \#387, GOES M1.0 class),
{\bf an event with multi-peak characteristics} (Fig.~2; \#219, GOES M2.0 class), 
and one of the largest events (Fig.~3; \#12, GOES X2.2 class). In all three
cases we show the time evolution of the most important fit 
parameters in the various panels (d through j) of Figs.~1-3: 
(d) the thermal emission measure $EM(t)$;
(e) the temperature evolution $T_e(t)$;
(f) the nonthermal photon flux $I_{\mathrm{nt}}(t)$ at 50 keV;
(g) the power law slope $\delta(t)$;
(h) the goodness-of-fit $\chi(t)$;
(i) the nonthermal power $P_{\mathrm{wt}}(t)$ using the low cutoff energy based on the
warm-target model (Section 2.3); and
(j) the low-energy cutoff $e_{\mathrm{wt}}(t)$ of the warm-target model.
{\bf In the examples shown in Figs.~1, 2 and 3 we see that the thermal 
emission measure $EM(t)$ increases during the rise time of the GOES flux, 
while the temperature $T_e(t)$ decreases, which indicates both density 
increases due to chromospheric evaporation as well as subsequent 
plasma cooling during the impulsive flare phase. Since multiple
heating and cooling cycles overlap during a flare, we see both effects
simultaneously. The cases shown in Figs.~1, 2 and 3 show also that
the nonthermal flux $I_{\mathrm{nt}}(t)$ (Figs.~1f, 2f, 3f) and the power
$P_{\mathrm{wt}}(t)$ (Figs.~1i, 2i, 3i) are correlated with the GOES
time derivative (Figs.~1a, 2a, 3a).} 

\subsection{	Goodness-of-Fit 	 			}

The goodness of the spectral fits computed with the OSPEX code is 
specified with the $\chi^2$-criterion, based on the least-square
difference between the theoretical spectral model (isothermal plus
power law nonthermal function) and the observed counts in the fitted
energy range [$\varepsilon_1$-$\varepsilon_2$]. The fitted energy
time interval (with a resolution of 1 keV) has about 
$n_{\mathrm{bin}} \approx 30-10 = 20$ energy bins, while the model 
has four ($n_{\mathrm{par}}=4)$ free parameters 
$(EM, T_e, A_{50}, \delta)$, yielding a degree of freedom
$n_{\mathrm{free}}=n_{\mathrm{bin}}-n_{\mathrm{par}} \approx 20-4 = 16$. 
In our spectral analysis of 191 flare events we performed spectral 
fits, with an average of $n_t \approx 27$ time steps per event, 
amounting to a total of $N_{\mathrm{spec}} \approx 
191 \times 27 = 5157$ spectral fits. The values $\chi(t)$ of three
events are shown in Figs.~1h, 2h, and 3h. The median values of these
three events are $\chi=1.4, 1.0$, and 1.3. We obtained in all 191
events a median goodness-of-fit value of $\chi < 2$,
after adjustment of the fitted energy range if necessary.
The mean and standard deviations of the median $\chi^2$-values of all
191 events is $\chi = 1.2 \pm 0.4$, which indicates that the
fitted spectral model is adequate in the chosen fitted energy range.
Of course, if one particular model,
such as the two-component thermal-nonthermal model chosen here
(Eq.~7), is found to be consistent with the data according to an
acceptable goodness-of-fit criterion, it does not rule out
alternative models. For instance, the thermal component is often
modeled with an iso-thermal (single-temperature) spectrum, while
a multi-thermal power law function was found to fit the thermal
flare component in most flares equally well (Aschwanden 2007).

\subsection{	Temperature Definitions 			}

A representative value for the electron temperature during a flare
can be defined in various ways. In paper II we measured the peak
temperature $T_{\mathrm{AIA}}$ of the differential emission measure (DEM)
distribution at the peak time of the flare, as well as the
emission measure-weighted temperature $T_w$ (Eq.~13 in Paper II),
which approximately characterizes the ``centroid'' of the
(logarithmic) DEM function. The mean ratio of these two temperature
values was found to be $q_T=T_{\mathrm{AIA}}/T_w=0.31$ within 
a standard deviation by a factor of $2.0$ (Fig.~4, left panel). The
emission measure-weighted temperature $T_w$ is generally found to be 
higher, because near-symmetric DEM functions as a function of 
the logarithmic temperature are highly asymmetric on a linear
temperature scale, with a centroid that is substantially higher than 
the logarithmic centroid. 

On the other hand, spectral fits of RHESSI data with an isothermal
component are known to have a strong bias towards the highest
temperatures occurring in a flare, because the fitted energy range
covers only the high-temperature tail of the DEM distribution function
(Battaglia et al.~2005; Ryan et al.~2014; Caspi et al.~2014).  
A statistical study demonstrated that the high-temperature bias of
RHESSI by fitting in the photon energy range of $\varepsilon \approx 
6-12$ keV amounts to a factor of $T_{R}/T_{\mathrm{AIA}}=1.9\pm1.0$
(Ryan et al.~2014).
Here we find that all RHESSI temperatures averaged during each flare
are found in a range of $T_{R}=16-40$ MK, which 
is about equal to the emission measure-weighted temperature, 
i.e., $T_{R}/T_w=0.90$ within a factor of 1.4 (Fig.~4, right panel). 
The 1-$\sigma$ ranges (containing 67\% of the values) of the various temperature 
definitions are $T_{\mathrm{AIA}} \approx 3-14$ MK, $T_w \approx 20-30$ MK, and
$T_{R} \approx 19-28$ MK.
So, we should keep these different temperature definitions in mind
when we calculate the low-energy cutoff $e_c(t)$ as a function of the
RHESSI temperature $T_{R}(t)$ (Eq.~8 for the warm-target model).

{\bf The most decisive parameter in the determination of the nonthermal energy
$E_{nt}$ is the low-energy cutoff $e_c$ (Eq.~4), which is directly proportional
to the temperature $T_e$ in the warm target (Eq.~8). What is the most likely
temperature of the warm target? The relevant temperature is a mixture of
pre-flare plasma temperatures and upflowing evaporating flare plasma.
In the absence of a sound model, we resort to the mean value of the
DEM peak temperatures determined in flaring active regions,
as determined with AIA in Paper II, yielding a mean value of $T_{AIA}=
8.8 \pm 6.0$ MK (Fig.~4 left panel), averaged over $N=380$ M and X-class
flare events. For the subset of 191 flare events observed with RHESSI,
this mean value is $T_e=8.6$ MK, or $k_B T_e = 0.74$ keV. Note that 
a deviation of the plasma temperature by a factor of two will result
into a deviation in the determination of the nonthermal energy $E_{nt}$ 
by about an order of magnitude (using a power law with a typical 
slope of $\gamma \approx 4$ in Eqs.~3 and 4).}

\subsection{		Nonthermal Energy Parameters			}

The nonthermal energy in electrons, calculated as a time integral
$E_{\mathrm{nt}}$ (Eq.~4), using the low-energy cutoff according to the
warm thick-target model $e_{\mathrm{wt}}(t)$ (Section 2.3; Eq.~8), 
or alternatively the thermal/nonther\-mal cross-over energy
$e_{\mathrm{co}}(t)$ (Section 2.2), is the main objective of this study. 
Examples of the time evolution of the
nonthermal parameters [$A(t), \delta(t)$, $e_{\mathrm{co}}(t)$, $e_{\mathrm{wt}}(t)$] 
and the resulting nonthermal energies $dE_{\mathrm{nt}}(t)$ are shown in 
Figs.~1-3. In Fig.~5 we show statistical results of
these parameters. Investigating the dependence of these parameters
on the flare temperature $T_{R}$ we find that both the
low-energy cutoff energy $e_{\mathrm{wt}}$ {\bf (Fig.~5a) as well as
the nonthermal (warm-target) energy $E_{\mathrm{nt}}$ (Fig.~5b) 
are {\bf uncorrelated} with the RHESSI temperature.
  
If we use the thermal-nonthermal cross-over method to estimate the 
low-energy cutoff, we find a systematically higher value, 
$e_{\mathrm{co}} \gapprox e_{\mathrm{wt}}$
(Fig.~5c). Consequently, the nonthermal energy
estimated with the cross-over method is systematically
lower than the nonthermal energy calculated with the warm-target model
(Fig.~5d). This result strongly depends on the assumption of the
warm-target temperature. Based on a mean temperature of $T_e=8.6$ MK 
found in the active regions analyzed here, 
we derive low-energy cutoff energies of $e_{\mathrm{wt}} =6.2\pm 1.6$ keV for 
the warm-target model, which is significantly lower than the cross-over
energies $e_{\mathrm{co}}=21 \pm 6$ keV. If we adopt the warm-target
model, we conclude that the cross-over method over-estimates the
low-energy cutoff and under-estimates the nonthermal energies.}

\subsection{	Comparison of Magnetic, Nonthermal, and Thermal Energies	}

In Fig.~6 we show scatterplots of the nonthermal energy $E_{\mathrm{nt}}$ 
measured here with other forms of previously determined energies, such as the
magnetic energy $E_{\mathrm{mag}}$ (Paper I) and the (total pre-impulsive and
post-impulsive) thermal energies $E_{\mathrm{th}}$ (Paper II). 
{\bf The energy ratios are characterized with the means of the logarithmic
energies in the following. The ratios between the three forms of energies
are shown separately for the cross-over method in the left-hand panels
of Fig.~6, and for the warm-target model in the right-hand panels of Fig.~6.

The ratios between the nonthermal energies and the magnetically dissipated 
energy is $E_{\mathrm{co}}/E_{\mathrm{mag}}=0.01$ for the cross-over method, 
or $E_{\mathrm{wt}}/E_{\mathrm{mag}}=0.41$ for the warm-target model,
respectively. Thus, the warm-target model yields ratios that are closer
to unity, which is expected in terms of magnetic reconnection processes,
where most of the magnetic energy is converted into particle acceleration. 
We find that the dissipated magnetic energy is sufficient to supply the
energy in nonthermal particles in 71\% for the warm-target model, or in
97\% for the cross-over model (Figs.~6a and 6b).

The ratios between the thermal energies and the magnetically dissipated 
energy is $E_{\mathrm{th}}/E_{\mathrm{mag}}=0.08$ for both the 
cross-over or the warm-target model (Fig.~6c and 6d).
We find that the dissipated magnetic energy is sufficient to supply the
thermal energy in 95\%.

Comparing the thermal with the nonthermal energies, we find a mean
a ratio of $E_{\mathrm{th}}/E_{\mathrm{wt}}=0.15$ for the warm-target
model, or $E_{\mathrm{th}}/E_{\mathrm{co}}=6.46$ for the cross-over
method. We find that the nonthermal energy is sufficient to supply the
thermal energy in 85\% for the warm-target model (Fig.~6f), but only in 
29\% for the cross-over method. Thus, the warm-target model yields
values that are closer to the expectations of the standard
thick-target model, where the thermal energy is entirely produced
by the nonthermal energy of precipitating (nonthermal) electrons.

We show the comparison of nonthermal and thermal energies also in form of
cumulative size distributions in Fig.~7, for the subset of 75 flare for
which all three forms of (magnetic, thermal, nonthermal) energies
could be calculated. We find that the nonthermal energy is typically
an order of magnitude larger than the thermal energy in the statistical
average. Then we find that the nonthermal energy is smaller than the
magnetic energy, as expected for magnetic reconnection processes, for
smaller flares with energies of $E_{nt} < 3 \times 10^{32}$ erg.
However, we find the opposite results for larger flares, with the
nonthermal energy exceeding the magnetically dissipated energy,
for large events with $E_{nt} > 3 \times 10^{32}$ erg.
Since the uncertainties in nonthermal energies are about an
order of magnitude and the dissipated magnetic energy exceeds
the nonthermal energy in 71\% (Fig.~6b), we suspect that the
largest nonthermal energies are over-estimated, which would
indicate that a higher value of the low-energy cutoff or a
higher flare plasma temperature (than the mean active region
temperature $T_e=8.6$ MK used here) could ameliorate the
over-estimated nonthermal energies.}

We compare the occurrence frequency distributions of magnetic,
nonthermal, and thermal energies, as well as those
of the direct RHESSI observables: the peak counts $P$, total counts $C$,
and durations $D$ (Fig.~8). As a caveat, we have to be aware
that these values for $P$ and $C$ are obtained
from the online RHESSI flare catalog, and thus are not
well-calibrated, as they do not take attenuation or decimation 
into account. Nevertheless, taking these raw values,
the magnetic and thermal energies have similar
power law slopes of $\alpha \approx 2.0$, while the nonthermal
energies have a slightly flatter slope of $\alpha_{\mathrm{nt}}=1.55 \pm 0.11$,
which can be compared with a previous study, where a power law slope
of $\alpha_{\mathrm{nt}}=1.53\pm0.02$ was found (Crosby et al.~1993). 
The latter study is actually
based on larger statistics, containing 2878 flare events observed
with HXRBS/SMM during 1980-1982 (Crosby et al.~1993), but with a higher
assumed low-energy cutoff of $e_c > 25$ keV.

\section{               DISCUSSION 				}

\subsection{	Energy Partition in Flares			} 

While we determined the dissipated magnetic energies $E_{\mathrm{mag}}$ (Paper I;
called $E_{\mathrm{diss}}$ therein), thermal energies $E_{\mathrm{th}}$ (Paper II), and 
the nonthermal energies $E_{\mathrm{nt}}$, we can ask now the question how the 
energy partition from primary to secondary energy dissipation works 
in solar flares. Many solar flare models are based on a magnetic 
reconnection process, where a stressed non-potential magnetic 
field becomes unstable and undergoes a reconfiguration towards a lower
magnetic energy state, releasing during this process some amount 
$E_{\mathrm{mag}}=q_{\mathrm{diss}} E_{\mathrm{free}}$ of the magnetic free energy $E_{\mathrm{free}}$ 
(defined by the difference between the non-potential and the potential 
energy, $E_{\mathrm{free}}=E_{\mathrm{np}}-E_p$). Excluding alternative energy sources,
we hypothesize that this dissipated magnetic energy $E_{\mathrm{mag}}$ is 
considered to be the entire available primary energy input, while
other energy conversion processes represent secondary steps that
need to add up in the energy budget, 
\begin{equation}
	E_{\mathrm{mag}} = ( E_{\mathrm{nt}} + E_{\mathrm{cme}} + ... ) > E_{\mathrm{nt}} \ ,
\end{equation}
such as the nonthermal energy $E_{\mathrm{nt}}$ that goes into acceleration
of particles, or the energy $E_{\mathrm{cme}}$ to accelerate an accompanying
coronal mass ejection (CME). The nonthermal energy $E_{\mathrm{nt}}$ may be further
subdivided into energies in electrons $E_{\mathrm{nt,e}}$ and ions $E_{\mathrm{nt,i}}$,
\begin{equation}
	E_{\mathrm{nt}} = ( E_{\mathrm{nt,e}} + E_{\mathrm{nt,i}} + ... ) > E_{\mathrm{nt,e}} \ ,
\end{equation}
while the CME energy $E_{\mathrm{cme}}$ consists of the
kinetic energy $E_{\mathrm{kin}}$ and the gravitational potential
energy $E_{\mathrm{grav}}$, and part of it may be converted into 
acceleration of particles in the interplanetary CME shock ($E_{\mathrm{nt,cme}}$),
which are particularly present in solar energetic particle (SEP) events,
\begin{equation}
	E_{\mathrm{cme}} = E_{\mathrm{kin}} + 
			   E_{\mathrm{grav}} + 
			   E_{\mathrm{nt,cme}} + ... \ .
\end{equation}
We have to be careful to avoid double-counting of secondary energies, 
because there may be some tertiary energy conversion processes, such
as heating of chromospheric plasma according to the thick-target 
bremsstrahlung model, $E_{\mathrm{th}}$, while upgoing nonthermal particles
escape into interplanetary space, carrying an energy of $E_{\mathrm{nt,esc}}$,
\begin{equation}
	E_{\mathrm{nt}} = ( E_{\mathrm{th}} + E_{\mathrm{nt,esc}} + ... ) > E_{\mathrm{th}} \ .
\end{equation}
Since we have measured only three types of energies so far,
$E_{\mathrm{mag}}, E_{\mathrm{nt}}$, and $E_{\mathrm{nt}}$, we can test only the inequalities given
on the righthand-side of Eqs.~(9) and (12) at this point.

Based on the nonthermal energies in electrons
determined in this work we can answer the question
whether the so far measured magnetic energy is sufficient
to accelerate the electrons 
observed in hard X-rays, i.e., $E_{\mathrm{mag}} > E_{\mathrm{nt}}$,
as expected for magnetic reconnection models.
Relying on the warm-target model we found that 41\% of the dissipated
magnetic energy (with a standard deviation of about an order of
magnitude) is converted into acceleration of nonthermal electrons,
or a total amount of $\approx 82\%$ for both electrons and ions in the
case of equipartition, while the rest is available to accelerate CMEs. 
There are few statistical estimates of the flare energy budget in literature 
(besides the work of Emslie et al.~2012; Warmuth and Mann 2016).
One early study quoted that the nonthermal energy in electrons $>20$ keV
contains 10-50\% of the total energy output for the August 1972 flares
(Lin and Hudson 1976; Hudson and Ryan 1995), which is consistent 
with our result of 41\% within the measurement uncertainties.

Comparing the energy ranges determined in this global flare energetics
project with those obtained from 38 events in Emslie et al.~(2012),
we find higher amounts of nonthermal flare electron energies 
in the statistical average,
covering the range of $E_{\mathrm{nt}} \approx (20-2000) \times 10^{30}$ erg 
(Fig.~9), which is mostly accounted for by a lower value of the
low-energy cutoff predicted by the 
warm-target model (Kontar et al.~2015) for some events, while
cutoff energies with the highest acceptable value of the 
$\chi^2$ was used in Emslie et al.~(2012).
The magnetically dissipated energies appear to be over-estimated by 
an order of magnitude (Fig.~9) in Emslie et al.~(2012), based on the ad hoc 
assumption that the dissipated energy amounts to 30\% of the potential 
field energy therein (Paper I). On the other hand, the thermal energies 
appear to be underestimated by at least an order of magnitude (Fig.~9)
in Emslie et al.~(2012) due to the isothermal approximation, 
as discussed in Paper II.

\subsection{Insufficiency of the Thick-Target Model?}

A second question we can answer is whether
the nonthermal energy in electrons is sufficient to
heat the flare plasma by the chromospheric evaporation process, 
as expected in the thick-target model according to the Neupert effect 
(Dennis and Zarro 1993), which requires $E_{\mathrm{nt}} > E_{\mathrm{th}}$. 
{\bf Based on the warm-target model we found a mean (logarithmic) ratio of 
$E_{\mathrm{th}}=0.15\ E_{\mathrm{nt}}$ (Fig.~6f). The fraction of flares
that have a thermal energy less than the nonthermal energy,
as expected in the standard thick-target model,
amounts in our analysis to $\approx 85\%$ for the warm-target method,
or $\approx 29\%$ for the cross-over model. 

This means that the thick-target model
could be insufficient to supply enough energy to explain the thermal
energy produced by the chromospheric evaporation process in about
15\% of the flares for the warm-target model, 
or in 71\% for the cross-over model.
Thus, the cross-over model would pose a series problem for the
thick-target model.}
The insufficiency of the thick-target model has been
addressed as a failure of the theoretical Neupert effect
(Veronig et al.~2005; Warmuth and Mann 2016),
which invokes testing of the correlation between the electron beam
power (from RHESSI) and the time derivative of the thermal energy
heating rate (from GOES). From such study it was concluded
that (1) fast electrons are {\sl not} the main source of soft X-ray
plasma supply and heating, (2) the beam low cutoff energy varies
with time, or (3) the theoretical Neupert effect is strongly
affected by the source geometry (Veronig et al.~2005). 
If the thermally dominated flares cannot be fully explained by the 
thick-target model, additional heating sources 
besides precipitating electrons would be required. The most popular 
alternative to the thick-target model is heating by thermal 
conduction fronts (Brown et al.~1979; Batchelor et al.~1985; 
Emslie and Brown 1980; Smith and Harmony 1982; Smith and Brown 1980;
Reep et al.~2016). Other forms of direct heating 
(for an overview see chapter 16 in Aschwanden 2004) occur via
(1) resistive or Joule heating processes, such as anomalous
resistivity heating (Duijveman et al.~1981; Holman 1985; Tsuneta
1985), ion-acoustic waves (Rosner et al.~1978b), electron
ion-cyclotron waves (Hinata 1980), (2) slow-shock heating 
(Cargill and Priest 1983; Hick and Priest 1989),
(3) electron beam heating by Coulomb collisional loss in the
corona (Fletcher 1995, 1996; Fletcher and Martens 1998),
(4) proton beam heating by kinetic Alfv\'en waves 
(Voitenko 1995, 1996), or (5) inductive current heating 
(Melrose 1995, 1997). 

{\bf The thick-target model fails to explain the observed amount of
thermal energy only in a small number of flares for the warm-target
model, while it is a larger number of events for the 
cross-over method. However, it is more likely that the cross-over 
method over-estimates the low-energy cutoff, which under-estimates
the nonthermal energies, while the physics-based warm-target model
leads to higher nonthermal energies, in which case the problem 
with the insufficiency of the thick-target model goes away.}

\subsection{	Nonthermal Low-Energy Cutoff in Flares		}		

We outlined two different methods to infer a low-energy
cutoff. The first method consists in measuring the cross-over
between the fitted thermal and nonthermal spectral components,
which yields an upper limit on the low-energy cutoff, 
but a statistical test demonstrates that the obtained values 
($e_{\mathrm{co}}=21 \pm 6$ keV) {\bf are significantly higher 
than those obtained from the warm-target model
($e_{\mathrm{wt}}=6.2 \pm 1.6$ keV).} 
There are pros and cons for each method. The cross-over method
requires a dominant thermal component, {\bf which is
not always detectable in the spectrum, in which case
the cross-over energy has a large uncertainty. The warm-target model
requires the measurement of the (warm) flare temperature, which
is measured at lower values from DEMs at EUV wavelengths than
from hard X-ray spectra observed with RHESSI.
Moreover, the spatial temperature 
distribution is very inhomogeneous and the location with the 
dominant temperature component relevant for the warm-target
collisional energy loss may be a mixture of colder pre-flare
plasma in active regions and heated evaporating flare plasma
at the location of instantaneous electron precipitation.
In summary, the value of the low-energy cutoff is strongly
dependent on the assumed warm-target temperature,
for which no physical model is established yet.}

In this study we investigated also the temporal evolution of the
low energy cutoff $e_c(t)$, for instance as shown in Fig.~1j, 2j, and 3j,
but we do not recognize a systematic pattern how the evolution
of this low-energy cutoff is related to other flare parameters. 

\section{               CONCLUSIONS                             }

The energy partition study of Emslie et al.~(2012) was restricted to 
38 large solar eruptive events (SEE). In a more comprehensive study
on the global flare energetics we choose a dataset that contains
the 400 largest (GOES M and X-class) flare events observed during 
the first 3.5 years of the SDO era. Previously we determined the
dissipated magnetic energies $E_{\mathrm{mag}}$ in these flares based on 
fitting the {\sl vertical-current approximation of a nonlinear force-free 
field (NLFFF)} solution to the loop geometries detected in EUV
images from SDO/AIA, a new method that could be applied to 177 events 
with a heliographic longitude of $\le 45^\circ$ (Paper I).
We also determined the thermal energy $E_{\mathrm{th}}$ in the soft X-ray
and EUV-emitting plasma during the flare peak times based on a
multi-temperature differential emission measure DEM forward-fitting
method to SDO/AIA image pixels with spatial synthesis,
which was applicable to 391 events 
(Paper II). In the present study we determined the nonthermal energy
$E_{\mathrm{nt}}$ contained in accelerated electrons based on spectral
fits to RHESSI data using the OSPEX software, which was applicable
to 191 events. The major conclusions of the new results emerging
from this study are: 

\begin{enumerate}
\item{The (logarithmic) mean energy ratio of the nonthermal
energy to the total magnetically dissipated flare energy is found to be
$E_{\mathrm{nt}}/E_{\mathrm{mag}}=0.41$, with a logarithmic standard 
deviation corresponding to a factor of $\approx 8$, which yields
an uncertainty $\sigma/\sqrt{N}=0.41/\sqrt{191}=0.03$ for
the mean, i.e., $E_{\mathrm{nt}}/E_{\mathrm{mag}}=0.41\pm0.03$.  
The majority ($\approx 85\%$) of the flare events fulfill the
inequality $E_{\mathrm{nt}}/E_{\mathrm{mag}} < 1$, which suggests 
that magnetic energy dissipation (most likely by a magnetic 
reconnection process) provides sufficient energy to accelerate 
the nonthermal electrons detected by bremsstrahlung in hard X-rays. 
Our results yield an order of magnitude higher electron acceleration 
efficiency than previous estimates, i.e., $E_{\mathrm{nt}}/E_{\mathrm{mag}}
=0.03\pm 0.005$ (with $N=37$, Emslie et al.~2012).}

\item{The (logarithmic) mean of the thermal energy $E_{\mathrm{th}}$ 
to the nonthermal energy $E_{\mathrm{nt}}$ is found to be {\bf
$E_{\mathrm{th}}/E_{\mathrm{nt}}=0.15$, with a logarithmic 
standard deviation corresponding to a factor of $\approx 7$.  
The fraction of flares with a thermal energy being
smaller than the nonthermal energy, as expected in the thick-target
model, is found to be the case for $\approx 85\%$ only. Therefore,
the thick-target model is sufficient to explain the full amount of
thermal energy in most flares, in the framework of the warm-target
model. The cross-over method shows the opposite tendency,
but we suspect that the cross-over method over-estimates the
low energy cutoff and under-estimates the nonthermal energies.
Previous estimates yielded a similar ratio, i.e., 
$E_{\mathrm{th}}/E_{\mathrm{nt}}=0.15$ (Emslie et al.~2012).}} 

\item{A corollary of the two previous conclusions is that the
thermal to magnetic energy ratio is $E_{\mathrm{th}}/E_{\mathrm{mag}}=0.08$. A total
of $95\%$ flares fulfils the inequality $E_{\mathrm{nt}}/E_{\mathrm{mag}} < 1$,
indicating that all thermal energy in flares is supplied by
magnetic energy. Previous estimates were a factor of 17 lower,
i.e., $E_{\mathrm{th}}/E_{\mathrm{mag}}=0.0045$ (Emslie et al.~2012), 
which would imply a very inefficient magnetic to thermal energy conversion 
process.}

\item{The largest uncertainty in the calculation of nonthermal
energies, the low-energy cutoff, is found to yield different values
for two used methods, i.e., $e_{\mathrm{wt}}=6.2 \pm 1.6$ keV
for the warm thick-target model, versus $e_{\mathrm{co}}=21 \pm 6$ keV 
for the thermal/nonthermal cross-over method. The calculation of
the nonthermal energies is highly sensitive to the value of the
low-energy cutoff, which strongly depends on the assumed 
(warm-target) temperature.} 

\item{The flare temperature can be characterized with three different
definitions, for which we found the following ($67\%$-standard deviation) 
ranges: $T_{\mathrm{AIA}} \approx 3-14$ MK for the AIA DEM peak temperature, 
$T_w \approx 20-30$ MK for the emission measure-weighted temperatures, and
$T_{R} \approx 17-36$ MK for the RHESSI high-temperature DEM tails. 
The median ratios are found to be $T_{\mathrm{AIA}}/T_w=0.31$ and
$T_{R}/T_w=0.90$. 
{\bf The mean active region temperature evaluated from DEMs with AIA,
$T_e=8.6$ MK, is used to estimate the low-energy cutoff $e_c$ of the 
nonthermal component according to the
warm-target model, i.e., $e_c \approx \delta (k_B T_{R})$. 
The low-energy cutoff $e_c$ of the nonthermal spectrum has 
a strong functional dependence on the temperature $T_{R}$.}}
\end{enumerate}

In summary, our measurements appear to confirm that the magnetically
dissipated energy is sufficient to explain thermal and nonthermal
energies in solar flares, which strongly supports the view that
magnetic reconnection processes are the primary energy source of flares. 
The nonthermal energy, which represents the primary energy source of the 
thick-target model, {\bf is sufficient to explain the full amount of
thermal energies in 71\% of the flares, according to the novel
warm-target model (Kontar et al.~2011). However, the derived
nonthermal energies are highly dependent on the
the assumed temperature in the warm-target plasma, for which
a sound physical model should be developed (see for instance 
Appendix A and B), before it becomes a useful 
tool to estimate the low-energy cutoff of nonthermal energy spectra.}
Future studies of this global flare energetics project
may also quantify additional forms of energies, such as the kinetic
energy in CMEs, and radiated energies in soft X-rays, EUV, and
white-light (bolometric luminosity).

\bigskip
\acknowledgements
We acknowledge useful comments from an anonymous referee
and discussions with Brian Dennis,
Gordon Emslie, Iain Hannah, Ryan Milligan, Linhui Sui, Daniel Ryan,
Richard Schwartz, Alexander Warmuth, and software support from 
Kim Tolbert and Samuel Freeland. 
This work was partially supported by NASA contract
NAS5-98033 of the RHESSI mission through University of California,
Berkeley (subcontract SA2241-26308PG), and 
by NASA contract NNG 04EA00C of the SDO/AIA instrument. 
AC and JMM were also supported by NASA grant NNX15AK26G.

\subsection*{ 	APPENDIX A : Collisional Time-of-Flight Model	}

We can derive a collisional time-of-flight  model for the thermal/non-thermal 
cross-over energy that is complementary to the warm-target model of Kontar 
et al.~(2015). For stochastic acceleration models, where particles 
gain and lose energy randomly, the collisional deflection time yields 
an upper time limit during which a particle can be efficiently accelerated. 
The balance between acceleration and collisions can lead to the 
formation of a kappa-distribution according to some solar flares models
(Bian et al.~2014). For solar flares, we can thus estimate the 
cross-over energy between collisional and collisionless electrons 
by setting the collisional deflection time $t_{\mathrm{defl}}$,
$$
        t_{\mathrm{defl}} \approx 0.95 \times 10^8
                \left( {e_{\mathrm{keV}}^{3/2} \over n_e} \right)
                \left( { 20 \over \ln \Lambda} \right) \ ,
	\eqno(A1)
$$
where $\ln \Lambda \approx 20$ is the Coulomb logarithm,
equal to the (relativistic) time-of-flight propagation time
between the coronal acceleration site and the chromospheric
thick-target energy loss site,
$$
        t_{\mathrm{TOF}} = {L_{\mathrm{TOF}} \over v} = {L_{\mathrm{TOF}} \over \beta c} \ ,
	\eqno(A2)
$$
where the relativistic speed $\beta = v/c$,
$$
        \beta = \sqrt{ 1  - {1 \over \gamma^2} } \ ,
	\eqno(A3)
$$
is related to the kinetic energy $e_{\mathrm{kin}}$ of the electron by
$$
        e_{\mathrm{kin}} = m_e c^2 (\gamma - 1) = 511\ (\gamma - 1) \ {\rm [keV]} \ ,
	\eqno(A4)
$$
where $\gamma$ represents here the relativistic Lorentz factor
(not to be confused with the spectral slope of the photon spectrum
used above, i.e., Eq.~1). So, setting these two time scales equal,
$$
        t_{\mathrm{defl}} = t_{\mathrm{TOF}} \ ,
	\eqno(A5)
$$
yields the relationship, using $\ln \Lambda \approx 20$,
$$
        (\gamma - 1)^{3/2} (1 - {1 \over \gamma^2} )^{1/2}
        = { L_{\mathrm{TOF}}\ n_e \over 0.95 \times 10^8 \times 511^{3/2}\ c } \ .
	\eqno(A6)
$$
Using the low-relativistic approximation (for $\gamma \gapprox 1$),
$$
        (\gamma - 1)^{3/2} (1 - {1 \over \gamma^2} )^{1/2} = 
        (\gamma - 1)^{3/2} {(\gamma-1)^{1/2} (\gamma+1)^{1/2} \over \gamma} =  
        {(\gamma - 1)^2 (\gamma+1)^{1/2} \over \gamma}  
        \approx (\gamma - 1)^2 \sqrt{2} \ ,
	\eqno(A7)
$$
we obtain,
$$
	(\gamma-1)^2 \ \sqrt{2} \approx 0.003 \times
        \left( {L_{\mathrm{TOF}} \over 10^9 \ {\rm cm}} \right)
        \left( {n_e \over 10^{11} \ {\rm cm}^{-3}} \right)
        \quad {\rm [keV]} \ .
	\eqno(A8)
$$
and by inserting $(\gamma-1) = e_{c}/511$ keV from Eq.~(A4) we find 
the cross-over energy $e_c \approx e_{\mathrm{kin}}$ 
can be explicitly expressed as
$$
        e_c \approx 24
        \left( {L_{\mathrm{TOF}} \over 10^9 \ {\rm cm}} \right)^{1/2}
        \left( {n_e \over 10^{11} \ {\rm cm}^{-3}} \right)^{1/2}
        \quad {\rm [keV]} \ .
	\eqno(A9)
$$
This expression requires the measurement of a mean length scale
$L_{\mathrm{TOF}}$ of flare loops and an average electron density $n_e$
where electrons propagate. 

Turning the argument around predicts a time-of-flight distance
$L_{\mathrm{TOF}} \propto e_c^2/n_e$ as a function of the low-energy cutoff
$e_c$, which is a similar concept that has been applied to model
the size $L$ of the acceleration region as a function of the electron
energy $e$, i.e., $(L-L_0) \propto e^2/n_e$ (Guo et al.~2012a,b; 2013;
Xu, Emslie, and Hurford 2008). 

\subsection*{ 	APPENDIX B : The Rosner-Tucker-Vaiana Model	}

At the peak time of a flare, an energy balance between plasma 
heating and cooling occurs at the turnover point of the temperature 
maximum (Aschwanden and Tsiklauri 2009), which corresponds
to the scaling law of Rosner, Tucker, and Vaiana (1978a) that was
originally applied to steady-state heating of coronal loops,
where an energy balance between the heating rate and the conductive
and radiative cooling time is assumed. The RTV scaling law,
$T^3 \propto p L$, can be expressed in terms of the ideal gas pressure
$p = 3 n_e k_B T$, which yields for the loop apex temperature $T_{\mathrm{RTV}}$,
$$
        T_{\mathrm{RTV}} = 0.0011 \ (n_e L_{\mathrm{RTV}})^{1/2} \ .
	\eqno(B1)
$$
The loop half length and time-of-flight distance scale approximately
with the flare size, $L_{\mathrm{TOF}} \approx L_{\mathrm{RTV}} \approx L$.
Interestingly, the parameter combination $(n_e L)^{1/2}$ occurs
also in the expression for the collisional low-energy cutoff
(Eq.~A9), so that we can insert the RTV scaling law and obtain
an expression for the low-energy cutoff energy $e_c$ that depends
on the temperature $T_{\mathrm{RTV}}$ only,
$$
        e_c \approx 25 \ (k_B T_{\mathrm{RTV}}) \quad {\rm [keV]} \ ,
	\eqno(B2)
$$
which is similar to the result of the warm-target model (Eq.~8).
However, while the warm-target model is applied to the evaporating
upflowing flare plasma, which has temperatures of $T_e \approx 10-25$ MK,
the collisional deflection model should be applied to the temperature
of the cooler preflare loops, where the accelerated particles propagate
from the acceleration site to the thick-target site. These cooler
preflare loops may have typical coronal temperatures of
$T_{\mathrm{RTV}} \approx 5-6$ MK ($\approx 0.43-0.52$ keV) in active
regions (Hara et al.~1992), which predicts
then low-energy cutoff energies of $e_c=11-13$ keV.
If the time-of-flight distance $L_{\mathrm{TOF}}$ is corrected for
magnetic twist and the pitch angle of the electrons, the effective
time-of-flight distance is about $L_{\mathrm{TOF}} \lapprox 2 L$
(Aschwanden et al.~1996), which
increases the low-energy cutoff energy by a factor of $\sqrt{2}$,
predicting values of $e_c=15-18$ keV. Combining Eqs.~(8) and (B2), the
RTV model predicts a relationship between the preflare temperature
$T_{\mathrm{pre}}=T_{\mathrm{RTV}}$ and the (maximum) 
flare temperature $T_{\mathrm{flare}}$,
$$
	T_{\mathrm{pre}} \approx T_{\mathrm{flare}} 
	\left( {\delta \over 25} \right) \ ,
	\eqno(B3)
$$
which yields $T_{\mathrm{pre}} \approx (0.12 - 0.24) \ T_{\mathrm{flare}}$
for a range of spectral slopes $\delta \approx 3-6$. Given the fact
that flare temperatures are typically found in the range of
$T_{flare} \approx 10-25$ MK, while preflare temperatures amount
to typical coronal temperatures in active regions, $T_{pre} \approx
1-4$ MK, we would expect indeed temperature ratios of
$T_{pre}/T_{flare} \approx 0.1-0.16$.

\subsection*{ 	APPENDIX C : The Runaway Acceleration Model }

Some particle acceleration models involve DC electric fields
that accelerate electrons and ions out of the bulk plasma.
Since the frictional drag on the electrons decreases with
increasing particle velocity ($\nu \propto v^{-3}$),
electrons in the initial thermal distribution with a high
enough velocity will not be confined to the bulk current,
but will be freely accelerated out of the thermal distribution
(Kuijpers et al.~1981; Holman 1985), 
a process that is called runaway acceleration.
A thermal electron of velocity $v_e$ will run away if the
electric field strength is greater than the Dreicer field $E_D$,
$$
        E_D = {m \over e} v_e \nu_e \ ,
	\eqno(C1)
$$
where $m$ is the electron mass, $e$ the electron charge,
$v_e$ the electron velocity, and $\nu_e$ the electron collision
frequency. Since the square of the (non-relativistic) speed $v_e$
scales with the kinetic energy, $E_{\mathrm{kin}} = (1/2) m_e v_e^2$,
the critical runaway energy $E_{\mathrm{ra}}$ can be characterized by the
ratio of the critical velocity $v_e$ to the thermal speed $v_{\mathrm{th}}$,
$$
        E_{\mathrm{ra}} = E_{\mathrm{th}} \left( { v_e \over v_{\mathrm{th}}} \right)^2 \ ,
	\eqno(C2)
$$
We can associate this critical runaway energy $E_{\mathrm{ra}}$
with the low-energy cutoff $e_c$ and obtain again a relationship
that scales with the plasma temperature $T_e$ for a given critical
velocity ratio,
$$
        e_c \approx E_{\mathrm{ra}} = k_B T_e \ \left( { v_e \over v_{\mathrm{th}}} \right)^2
        \quad {\rm [keV]} \ .
	\eqno(C3)
$$
Thus, for a typical velocity ratio of $(v_e/v_{\mathrm{th}}) \approx$ 2$-$3
and a plasma temperature range of $T_e \approx 5-6$ MK
$\approx 0.43-0.52$ keV in active regions, this model predicts a
range of $e_c \approx 1.7-8.3$ keV. Combining the relationships
of the warm-target model (Eq.~8) and the runaway acceleration model
(Eq.~C1) yields then a prediction for the nonthermal speed ratio
of the runaway electrons,
$$
	\left( {v_e \over v_{\mathrm{th}}} \right) \approx \sqrt{\delta}
	\approx (1.7 - 2.4) \ ,
	\eqno(C4)
$$
which is consistent with solar parameters used in runway models
(Kuijpers et al.~1981; Holman 1985). Implications of runway
acceleration models for sub-Dreicer and super-Dreicer fields
are discussed also in Guo, Emslie and Piana (2013) and 
Miller et al. (1997).


\clearpage

\begin{deluxetable}{rrrrrrrrrrrrr}
\tabletypesize{\footnotesize}
\tablecaption{Nonthermal energy parameters derived in 191 flare events 
observed with RHESSI: The soft X-ray flare duration $d$ (column 5), 
the peak counts $P$ (column 6), the total counts $C$ (column 7), 
the fitted energy range (column 8), the (warm-target) lower cutoff energy
$e_{wt}$ for a mean temperature of $T_e=8.6$ MK in flaring active regions
(column 9), the (warm-target) nonthermal energy $E_{\mathrm{wt}}$
(Column 10), the ratio of the thermal energy $E_{\mathrm{th}}$ 
to the (warm-target) nonthermal energy $E_{\mathrm{wt}}$ (column 11),
and the ratio of the the (warm-target) nonthermal energy $E_{\mathrm{wt}}$ 
to the magnetic energy $E_{\mathrm{mag}}$ (column 12). Questionable solar 
flare events, detected in the front detectors without position, are 
flagged with a $(^*)$ sign (in column 11).}
\tablewidth{0pt}
\tablehead{
\colhead{\#}&
\colhead{Flare}&
\colhead{GOES}&
\colhead{Helio-}&
\colhead{Flare}&
\colhead{Peak}&
\colhead{Total}&
\colhead{Fitted}&
\colhead{Cutoff}&
\colhead{Nonthermal}&
\colhead{Energy}&
\colhead{Energy}\\
\colhead{}&
\colhead{start time}&
\colhead{class}&
\colhead{graphic}&
\colhead{duration}&
\colhead{counts}&
\colhead{counts}&
\colhead{energy}&
\colhead{energy}&
\colhead{energy}&
\colhead{ratio}&
\colhead{ratio}\\
\colhead{}&
\colhead{}&
\colhead{}&
\colhead{position}&
\colhead{$d$}&
\colhead{$P$}&
\colhead{$C$}&
\colhead{range}&
\colhead{$e_{\mathrm{wt}}$}&
\colhead{$E_{\mathrm{wt}}$}&
\colhead{$E_{\mathrm{th}}/E_{\mathrm{wt}}$}&
\colhead{$E_{\mathrm{wt}}/E_{\mathrm{mag}}$}\\
\colhead{}&
\colhead{}&
\colhead{}&
\colhead{}&
\colhead{(s)}&
\colhead{(cts/s)}&
\colhead{(cts)}&
\colhead{(keV)}&
\colhead{(keV)}&
\colhead{(erg)}&
\colhead{}&
\colhead{}\\}
\startdata
   1 & 20100612 0030 & M2.0 & N23W47 &   904 &     92 &    1.3E+05 & [ 8-20] &  2.6 &    1.0E+30 &     6.98$^*$ & ... \\ 
   2 & 20100613 0530 & M1.0 & S24W82 &  1852 &    688 &    1.3E+06 & [ 6-20] &  4.9 &    5.4E+28 &    75.95$^*$ & ... \\ 
   4 & 20101016 1907 & M2.9 & S18W26 &  1572 &   3312 &    3.1E+06 & [ 6-26] &  5.6 &    6.2E+31 &     0.31 &     2.21 \\ 
   6 & 20101105 1243 & M1.0 & S20E75 &  2980 &    400 &    2.2E+06 & [ 6-20] &  7.1 &    6.5E+31 &     0.12 & ... \\ 
   8 & 20110128 0044 & M1.3 & N16W88 &  1760 &   1968 &    4.8E+06 & [ 6-20] &  7.2 &    3.9E+31 & ... & ... \\ 
  10 & 20110213 1728 & M6.6 & S21E04 &  2324 &   6384 &    2.5E+07 & [ 8-30] &  8.3 &    9.3E+32 &    0.022$^*$ &    10.98 \\ 
  12 & 20110215 0144 & X2.2 & S21W12 &  2628 &  26868 &    9.8E+07 & [10-50] &  5.8 &    1.1E+33 &    0.073 &     9.34 \\ 
  13 & 20110216 0132 & M1.0 & S22W27 &  1368 &   1072 &    1.5E+06 & [ 8-40] &  6.8 &    4.4E+31 &     0.17 &     0.39 \\ 
  15 & 20110216 1419 & M1.6 & S23W33 &  1692 &   1039 &    1.3E+06 & [ 6-30] &  6.9 &    3.8E+31 &     0.17 &     0.21 \\ 
  16 & 20110218 0955 & M6.6 & S21W55 &  1780 &   6082 &    6.5E+06 & [ 6-30] &  6.3 &    5.3E+32 &   0.0080 &    38.49 \\ 
  18 & 20110218 1259 & M1.4 & S20W70 &  1944 &   1904 &    3.6E+06 & [ 6-30] &  6.1 &    2.4E+31 &    0.088 & ... \\ 
  19 & 20110218 1400 & M1.0 & N17E04 &  1264 &    432 &    6.4E+05 & [ 8-20] &  6.8 &    1.0E+31 &     0.49 &     0.39 \\ 
  20 & 20110218 2056 & M1.3 & N15E00 &   884 &   1200 &    2.2E+06 & [ 6-30] &  7.0 &    4.7E+31 &    0.095 &     3.11 \\ 
  21 & 20110224 0723 & M3.5 & N14E87 &  3332 &   2032 &    5.0E+06 & [ 8-30] &  4.8 &    2.9E+31 &     0.58 & ... \\ 
  22 & 20110228 1238 & M1.1 & N22E35 &   732 &    688 &    1.2E+06 & [10-30] &  6.3 &    8.6E+31 &    0.074 &     2.88 \\ 
  23 & 20110307 0500 & M1.2 & N23W47 &  1340 &    880 &    1.5E+06 & [ 6-30] &  7.4 &    2.4E+31 &    0.019 & ... \\ 
  26 & 20110307 0914 & M1.8 & N27W46 &   348 &   1776 &    1.6E+06 & [ 6-30] &  4.4 &    1.8E+31 &   0.0093 & ... \\ 
  28 & 20110307 1943 & M3.7 & N30W48 &  3196 &   1328 &    6.7E+06 & [10-30] &  3.4 &    1.8E+31 &     1.30 & ... \\ 
  29 & 20110307 2145 & M1.5 & S17W82 &  1232 &    720 &    1.0E+06 & [ 8-30] &  6.4 &    4.0E+31 &    0.038 & ... \\ 
  30 & 20110308 0224 & M1.3 & S18W80 &  1460 &    752 &    6.8E+05 & [ 6-30] &  7.3 &    3.2E+31 &    0.088 & ... \\ 
  31 & 20110308 0337 & M1.5 & S21E72 &  2768 &    108 &    6.5E+05 & [12-30] &  3.9 &    2.8E+30 &     4.80 & ... \\ 
  33 & 20110308 1808 & M4.4 & S17W88 &   848 &   1712 &    5.6E+06 & [ 8-30] &  6.3 &    7.8E+32 &    0.020 & ... \\ 
  34 & 20110308 1946 & M1.5 & S19W87 &  6044 &    176 &    1.3E+06 & [ 6-20] &  6.3 &    3.7E+31 &     0.17 & ... \\ 
  37 & 20110309 2313 & X1.5 & N10W11 &  1660 &   4938 &    8.3E+06 & [10-40] &  5.8 &    1.1E+33 &    0.074 &     4.25 \\ 
  38 & 20110310 2234 & M1.1 & S25W86 &  1588 &    192 &    3.1E+05 & [ 8-30] &  6.7 &    4.5E+31 &    0.016 &     0.16 \\ 
  40 & 20110314 1930 & M4.2 & N16W49 &  2308 &   2988 &    3.3E+06 & [ 8-30] &  8.2 &    4.1E+32 &    0.021 & ... \\ 
  41 & 20110315 0018 & M1.0 & N11W83 &  1500 &   1648 &    7.1E+05 & [ 8-30] &  5.0 &    4.6E+30 &    0.077 & ... \\ 
  46 & 20110422 0435 & M1.8 & S19E40 &  3124 &    880 &    3.5E+06 & [10-30] &  6.7 &    1.1E+32 &    0.098 &     2.47 \\ 
  48 & 20110528 2109 & M1.1 & S21E70 &  2848 &    624 &    2.1E+06 & [ 6-30] &  7.3 &    1.4E+31 &     0.39 & ... \\ 
  49 & 20110529 1008 & M1.4 & S20E64 &  3552 &    448 &    3.5E+06 & [ 7-25] &  6.5 &    5.3E+31 &     0.15 & ... \\ 
  50 & 20110607 0616 & M2.5 & S22W53 &  3608 &    944 &    5.1E+06 & [ 8-30] &  3.3 &    1.4E+31 &     1.92 & ... \\ 
  51 & 20110614 2136 & M1.3 & N14E77 &  2356 &    688 &    1.7E+06 & [ 6-30] &  5.3 &    5.5E+31 &     0.13 & ... \\ 
  52 & 20110727 1548 & M1.1 & N20E41 &  2004 &    256 &    4.6E+05 & [ 6-30] &  6.8 &    6.4E+30 &     1.86 &     0.20 \\ 
  53 & 20110730 0204 & M9.3 & N16E35 &  1460 &   6115 &    6.5E+06 & [ 8-30] &  6.9 &    1.0E+33 &    0.028 &    11.05 \\ 
  54 & 20110802 0519 & M1.4 & N16W11 &  6208 &   1895 &    3.3E+06 & [10-30] &  5.3 &    1.1E+31 &     0.97 &    0.096 \\ 
  55 & 20110803 0308 & M1.1 & N15W23 &  2760 &    944 &    2.4E+06 & [ 6-30] &  6.9 &    3.4E+31 &     0.12 &     1.61 \\ 
  56 & 20110803 0429 & M1.7 & N16E10 &  1268 &   2160 &    1.6E+06 & [ 8-30] &  6.0 &    3.2E+31 &    0.098 &     0.14 \\ 
  61 & 20110809 0748 & X6.9 & N20W69 &  2256 &  53158 &    7.3E+07 & [12-40] &  5.5 &    3.2E+33 &    0.041 & ... \\ 
  63 & 20110905 0408 & M1.6 & N18W87 &  1516 &    624 &    2.3E+06 & [ 6-30] &  6.7 &    1.5E+31 &     0.18 & ... \\ 
  64 & 20110905 0727 & M1.2 & N18W87 &  2464 &    624 &    2.0E+06 & [10-25] & 11.7 &    3.5E+29 &     3.44 & ... \\ 
  65 & 20110906 0135 & M5.3 & N15W03 &   692 &   4724 &    3.9E+06 & [10-40] &  6.8 &    3.2E+32 &    0.069 &     2.86 \\ 
  66 & 20110906 2212 & X2.1 & N16W15 &  1024 &  21072 &    2.3E+07 & [12-40] &  5.0 &    7.6E+31 &     0.68 &     0.41 \\ 
  68 & 20110908 1532 & M6.7 & N17W39 &  1764 &   2439 &    4.7E+06 & [ 8-25] &  7.3 &    1.5E+33 &    0.019 &    10.99 \\ 
  69 & 20110909 0601 & M2.7 & N14W48 &  1644 &   3824 &    6.3E+06 & [10-40] &  5.2 &    8.9E+31 &     0.20 & ... \\ 
  70 & 20110909 1239 & M1.2 & N15W50 &   408 &     96 &    1.0E+05 & [ 7-30] &  5.8 &    9.2E+30 &     0.41 & ... \\ 
  71 & 20110910 0718 & M1.1 & N14W64 &  2488 &    688 &    3.0E+06 & [10-30] &  7.3 &    4.1E+31 &     0.14 & ... \\ 
  73 & 20110922 0953 & M1.1 & N24W55 &  1508 &    624 &    1.3E+06 & [ 9-30] &  8.2 &    4.2E+31 &    0.084 & ... \\ 
  75 & 20110923 0147 & M1.6 & N24W64 &  1832 &    624 &    2.1E+06 & [10-30] &  8.7 &    4.0E+31 &    0.093 & ... \\ 
  76 & 20110923 2154 & M1.6 & N12E56 &  2456 &   5616 &    2.2E+06 & [10-30] &  8.3 &    4.7E+31 &     0.14 & ... \\ 
  77 & 20110923 2348 & M1.9 & N12E56 &  2020 &   1008 &    2.7E+06 & [ 8-30] &  5.6 &    7.1E+31 &     0.20 & ... \\ 
  78 & 20110924 0921 & X1.9 & N13E61 &  3008 &  18653 &    4.4E+07 & [ 8-50] &  7.4 &    8.2E+33 &   0.0027 & ... \\ 
  81 & 20110924 1719 & M3.1 & N13E54 &  1324 &   2160 &    3.3E+06 & [ 6-30] &  5.2 &    1.2E+32 &    0.028 & ... \\ 
  83 & 20110924 1909 & M3.0 & N15E50 &  1068 &   1520 &    4.0E+06 & [ 7-30] &  5.4 &    1.1E+32 &     0.22 & ... \\ 
  84 & 20110924 2029 & M5.8 & N13E52 &  1180 &   5051 &    8.1E+06 & [ 8-40] &  6.7 &    2.1E+32 &    0.042 & ... \\ 
  86 & 20110924 2345 & M1.0 & S28W66 &  2596 &    336 &    1.3E+06 & [10-30] &  4.3 &    2.9E+30 &     0.53 & ... \\ 
  88 & 20110925 0431 & M7.4 & N13E50 &  3640 &   5462 &    2.7E+07 & [ 9-30] &  6.9 &    2.1E+33 &    0.018 & ... \\ 
  90 & 20110925 0925 & M1.5 & S28W71 &  2720 &    656 &    2.7E+06 & [ 7-30] &  6.9 &    5.2E+31 &    0.074 & ... \\ 
  91 & 20110925 1526 & M3.7 & N15E39 &   676 &   1840 &    2.5E+06 & [ 7-30] &  6.5 &    2.7E+31 &     0.64 &    0.058 \\ 
  93 & 20110926 0506 & M4.0 & N15E35 &   572 &   1957 &    2.5E+06 & [10-30] &  7.4 &    3.6E+32 &    0.032 &     0.51 \\ 
  98 & 20111002 0037 & M3.9 & N10W13 &  3696 &   4336 &    9.4E+06 & [10-30] &  6.9 &    4.2E+32 &    0.044 &     6.62 \\ 
 100 & 20111020 0310 & M1.6 & N18W88 &  1044 &   1392 &    3.5E+06 & [10-30] &  7.2 &    1.5E+32 &    0.012 & ... \\ 
 101 & 20111021 1253 & M1.3 & N05W79 &   760 &    624 &    9.9E+05 & [ 6-30] &  5.3 &    9.3E+30 &     0.28 & ... \\ 
 103 & 20111031 1455 & M1.1 & N20E88 &  3980 &   1392 &    3.8E+06 & [10-30] &  6.7 &    1.3E+32 &   0.0070 & ... \\ 
 110 & 20111105 0308 & M3.7 & N20E47 &  3752 &   1136 &    9.1E+06 & [10-30] &  7.9 &    1.0E+32 &     0.13 & ... \\ 
 111 & 20111105 1110 & M1.1 & N22E43 &  2392 &    320 &    9.7E+05 & [10-30] &  6.9 &    1.3E+31 &     0.25 &    0.044 \\ 
 116 & 20111115 0903 & M1.2 & N21W72 &  2448 &    656 &    1.6E+06 & [ 8-30] &  6.2 &    2.3E+31 &     0.12 & ... \\ 
 120 & 20111226 0213 & M1.5 & S18W34 &  2812 &    624 &    1.4E+06 & [10-30] &  5.6 &    6.8E+30 &     1.21 &     0.72 \\ 
 121 & 20111226 2012 & M2.3 & S18W44 &  1512 &   1456 &    3.2E+06 & [10-30] &  6.7 &    1.0E+32 & ... &     3.98 \\ 
 122 & 20111229 1340 & M1.9 & S25E70 &  2368 &    848 &    1.6E+06 & [10-30] &  7.4 &    2.9E+31 &     0.35 & ... \\ 
 123 & 20111229 2143 & M2.0 & S25E67 &   632 &   1008 &    1.2E+06 & [10-30] &  7.7 &    8.4E+31 &    0.079 & ... \\ 
 125 & 20111231 1309 & M2.4 & S25E46 &  1892 &   1584 &    1.6E+06 & [10-30] &  6.7 &    8.3E+31 &    0.049 & ... \\ 
 126 & 20111231 1616 & M1.5 & S22E42 &  1272 &    656 &    9.2E+05 & [10-30] &  7.1 &    4.6E+31 &     0.18 &     0.28 \\ 
 154 & 20120317 2032 & M1.3 & S25W28 &  1236 &   1136 &    8.2E+05 & [10-25] &  7.3 &    1.8E+31 &     0.35 &     0.65 \\ 
 156 & 20120416 1724 & M1.7 & N14E88 &  1932 &    352 &    1.5E+06 & [10-20] &  7.5 &    4.0E+31 &     0.37 & ... \\ 
 157 & 20120427 0815 & M1.0 & N13W26 &   732 &    528 &    6.4E+05 & [10-30] &  6.2 &    2.1E+31 &     0.34 &     4.72 \\ 
 158 & 20120505 1319 & M1.4 & N11E78 &   200 &    560 &    1.4E+05 & [10-30] &  1.6 &    5.5E+30 &     0.71$^*$ & ... \\ 
 159 & 20120505 2256 & M1.3 & N11E73 &   624 &   1200 &    9.6E+05 & [10-30] &  5.8 &    3.8E+31 &    0.091 & ... \\ 
 160 & 20120506 0112 & M1.1 & N11E73 &  1684 &    976 &    6.7E+05 & [10-30] &  5.7 &    1.2E+31 &     0.16 & ... \\ 
 163 & 20120508 1302 & M1.4 & N13E46 &   432 &   1264 &    1.1E+06 & [10-30] &  4.9 &    1.9E+31 &     0.25 & ... \\ 
 167 & 20120510 0411 & M5.7 & N12E19 &  1128 &   3339 &    5.9E+06 & [10-30] &  3.1 &    2.5E+30 &     7.59 &    0.017 \\ 
 168 & 20120510 2020 & M1.7 & N12E10 &  1612 &   1712 &    2.3E+06 & [10-30] &  5.4 &    6.4E+31 &     0.17 &     0.50 \\ 
 169 & 20120517 0125 & M5.1 & N07W88 &  2708 &   2416 &    1.3E+07 & [10-30] &  4.7 &    4.1E+31 &     0.96 & ... \\ 
 170 & 20120603 1748 & M3.3 & N15E33 &   852 &   1648 &    1.3E+06 & [10-30] &  4.2 &    9.0E+29 &    25.04 &    0.020 \\ 
 173 & 20120609 1645 & M1.8 & S16E76 &  1724 &   1264 &    1.8E+06 & [10-30] &  6.7 &    6.6E+31 &    0.047 & ... \\ 
 176 & 20120614 1252 & M1.9 & S19E06 &  9628 &    880 &    4.3E+06 & [10-30] &  3.8 &    2.6E+30 &     1.05$^*$ &    0.008 \\ 
 178 & 20120629 0913 & M2.2 & N15E37 &   696 &   2160 &    1.2E+06 & [10-30] &  6.5 &    2.3E+31 &     0.16 &     0.23 \\ 
 182 & 20120702 0026 & M1.1 & N15E01 &  1356 &    944 &    1.1E+06 & [10-30] &  6.4 &    1.5E+31 &     0.29 &     0.23 \\ 
 187 & 20120704 0947 & M5.3 & S17W18 &  2416 &   8339 &    9.5E+06 & [10-30] &  6.3 &    3.7E+32 &    0.020 &     2.24 \\ 
 189 & 20120704 1435 & M1.3 & S18W20 &   428 &    320 &    2.7E+05 & [10-25] &  2.5 &    2.6E+29 &    11.36 &    0.005 \\ 
 190 & 20120704 1633 & M1.8 & N14W33 &   828 &    192 &    3.6E+05 & [10-25] &  3.2 &    4.8E+29 &    35.19 &    0.017 \\ 
 195 & 20120705 0325 & M4.7 & S18W29 &  1768 &   4447 &    8.0E+06 & [10-30] &  6.6 &    3.5E+32 &    0.017 &     2.09 \\ 
 196 & 20120705 0649 & M1.1 & S17W29 &  1208 &    912 &    2.5E+06 & [10-30] &  6.7 &    5.3E+31 &    0.068 &     0.42 \\ 
 199 & 20120705 1139 & M6.1 & S18W32 &  1056 &   1536 &    1.9E+06 & [10-30] &  4.4 &    1.8E+31 &     1.12 &     0.14 \\ 
 200 & 20120705 1305 & M1.2 & S18W36 &  1400 &     80 &    2.8E+05 & [10-20] &  1.6 &    2.9E+29 &    30.20 &    0.002 \\ 
 203 & 20120706 0137 & M2.9 & S18W43 &  2748 &   4300 &    3.7E+06 & [10-30] &  5.2 &    3.8E+31 &     0.11 &     0.53 \\ 
 205 & 20120706 0817 & M1.5 & S12W48 &  1392 &   1392 &    1.8E+06 & [10-30] &  6.4 &    4.3E+31 &    0.060 & ... \\ 
 208 & 20120706 1848 & M1.3 & S15E88 &  1348 &    256 &    4.0E+05 & [10-30] &  5.7 &    3.2E+31 &     0.12 & ... \\ 
 210 & 20120707 0310 & M1.2 & S17W55 &  1664 &   1200 &    1.7E+06 & [10-30] &  6.6 &    5.5E+31 &    0.062 & ... \\ 
 211 & 20120707 0818 & M1.0 & S16E76 &   684 &    400 &    8.1E+05 & [10-30] &  4.8 &    5.1E+29 &     2.97 & ... \\ 
 212 & 20120707 1057 & M2.6 & S17W59 &   520 &   3065 &    3.5E+06 & [10-30] &  5.4 &    2.0E+32 &    0.025 & ... \\ 
 214 & 20120708 0944 & M1.1 & S16W70 &   768 &    784 &    8.7E+05 & [10-30] &  7.5 &    1.8E+31 &     0.15 & ... \\ 
 215 & 20120708 1206 & M1.4 & S16W72 &   160 &   1712 &    7.9E+05 & [10-30] &  5.6 &    3.4E+31 &    0.056 & ... \\ 
 219 & 20120710 0605 & M2.0 & S16E30 &  1848 &   1456 &    5.0E+06 & [10-30] &  8.1 &    1.2E+32 &    0.038 &     0.15 \\ 
 222 & 20120717 1203 & M1.7 & S20W88 & 20740 &    288 &    6.9E+06 & [10-25] & 10.5 &    1.3E+31 &     0.72 & ... \\ 
 223 & 20120719 0417 & M7.7 & S20W88 &  8532 &   3696 &    3.0E+07 & [10-25] &  5.8 &    2.5E+32 &    0.072 & ... \\ 
 228 & 20120806 0433 & M1.6 & S14E88 &   728 &   1264 &    1.3E+06 & [10-30] &  5.0 &    9.1E+30 &    0.029 &     0.70 \\ 
 230 & 20120817 1312 & M2.4 & N18E88 &  1512 &   2544 &    2.8E+06 & [10-30] &  5.9 &    5.6E+31 &    0.021 & ... \\ 
 235 & 20120818 2246 & M1.0 & N18E88 &  1036 &    400 &    7.8E+05 & [10-25] &  8.9 &    1.7E+31 &     0.28 & ... \\ 
 238 & 20120906 0406 & M1.6 & N04W61 &  2184 &   1456 &    2.0E+06 & [10-30] &  5.8 &    3.3E+31 &     0.16 & ... \\ 
 241 & 20120930 0427 & M1.3 & N12W81 &  2228 &   1072 &    2.1E+06 & [10-30] &  5.9 &    3.8E+31 &   0.0073 & ... \\ 
 245 & 20121020 1805 & M9.0 & S12E88 &  2116 &  12304 &    2.0E+07 & [10-30] &  6.1 &    8.6E+32 &   0.0089 & ... \\ 
 246 & 20121021 1946 & M1.3 & S13E78 &  2124 &    976 &    2.7E+06 & [10-30] &  7.1 &    9.3E+31 &    0.060 & ... \\ 
 248 & 20121023 0313 & X1.8 & S13E58 &  1380 &  16543 &    2.9E+07 & [10-25] &  7.0 &    2.5E+33 &   0.0046 & ... \\ 
 251 & 20121112 2313 & M2.0 & S25E48 &  2124 &   1840 &    3.2E+06 & [10-30] &  6.8 &    9.1E+31 &    0.044 & ... \\ 
 253 & 20121113 0542 & M2.5 & S26E44 &  1396 &   2288 &    3.1E+06 & [10-30] &  6.6 &    9.1E+31 &    0.072 &     0.83 \\ 
 255 & 20121114 0359 & M1.1 & S23E27 &  1352 &    720 &    6.3E+05 & [10-30] &  3.9 &    2.8E+29 &     6.36 &    0.007 \\ 
 256 & 20121120 1236 & M1.7 & N10E22 &   840 &   1200 &    9.7E+05 & [10-30] &  3.5 &    1.0E+30 &     0.15 &    0.048 \\ 
 257 & 20121120 1921 & M1.6 & N10E19 &   372 &   1072 &    5.4E+05 & [10-30] &  4.9 &    2.4E+30 &     1.45 &    0.077 \\ 
 258 & 20121121 0645 & M1.4 & N10E12 &   932 &   1136 &    2.0E+06 & [10-30] &  5.4 &    2.5E+31 &     0.25 &     0.43 \\ 
 261 & 20121127 2105 & M1.0 & S13W42 &  1668 &    720 &    9.2E+05 & [10-30] &  7.3 &    3.0E+31 &    0.075 &     0.71 \\ 
 262 & 20121128 2120 & M2.2 & S12W56 &  3044 &   1776 &    4.3E+06 & [10-30] &  6.6 &    6.8E+31 &     0.18 & ... \\ 
 264 & 20130111 0843 & M1.2 & N05E42 &  1180 &    880 &    2.0E+06 & [10-25] &  7.0 &    4.8E+31 &    0.066 &     0.24 \\ 
 266 & 20130113 0045 & M1.0 & N18W15 &   764 &   1264 &    6.6E+05 & [10-30] &  5.4 &    1.1E+31 &     0.17$^*$ &     0.53 \\ 
 268 & 20130217 1545 & M1.9 & N12E23 &   620 &   3312 &    1.5E+06 & [10-30] &  6.2 &    8.2E+30 &     0.12 &     0.45 \\ 
 271 & 20130321 2142 & M1.6 & N09W88 &  3516 &    560 &    3.7E+06 & [10-30] &  4.3 &    1.7E+31 &     0.50 & ... \\ 
 273 & 20130411 0655 & M6.5 & N11E13 &  1076 &   2160 &    2.8E+06 & [10-25] &  4.9 &    2.1E+31 &     1.90 &     0.42 \\ 
 274 & 20130412 1952 & M3.3 & N21W47 &  2012 &   2928 &    6.5E+06 & [10-30] &  6.5 &    1.5E+32 &    0.094 & ... \\ 
 276 & 20130502 0458 & M1.1 & N10W19 &  2380 &    448 &    1.3E+06 & [10-30] &  3.1 &    5.6E+29 &     7.36 &    0.009 \\ 
 277 & 20130503 1639 & M1.3 & N11W38 &  2872 &    649 &    2.3E+05 & [10-30] &  4.5 &    2.3E+30 &     0.37 &     0.16 \\ 
 278 & 20130503 1724 & M5.7 & N15E83 &  1316 &   3696 &    1.2E+07 & [10-30] &  6.1 &    2.7E+32 &    0.061 & ... \\ 
 283 & 20130512 2237 & M1.2 & N10E89 &  1872 &   1067 &    4.4E+06 & [10-30] &  7.2 &    2.4E+31 &     0.18 & ... \\ 
 284 & 20130513 0153 & X1.7 & N11E89 &  2496 &  13151 &    8.2E+07 & [10-30] &  6.3 &    6.8E+33 &   0.0033 & ... \\ 
 285 & 20130513 1157 & M1.3 & N10E89 &  1048 &   1264 &    1.3E+06 & [10-30] &  6.6 &    7.6E+31 &    0.012 & ... \\ 
 286 & 20130513 1548 & X2.8 & N08E89 &  1032 &  33601 &    7.3E+07 & [12-50] &  3.3 &    1.1E+31 &     6.19 & ... \\ 
 288 & 20130515 0125 & X1.2 & N10E68 &  3524 &   8656 &    3.9E+07 & [10-25] &  6.4 &    1.5E+33 &    0.026 & ... \\ 
 289 & 20130516 2136 & M1.3 & N11E40 &  1280 &    624 &    1.5E+06 & [10-30] &  9.8 &    3.7E+30 &     0.96 &     0.17 \\ 
 291 & 20130520 0516 & M1.7 & N09E89 &  1380 &    592 &    1.6E+06 & [10-25] &  6.8 &    3.6E+31 &    0.096 & ... \\ 
 292 & 20130522 1308 & M5.0 & N14W87 &  3248 &   1328 &    1.1E+07 & [10-30] &  4.4 &    1.3E+31 &     1.65 & ... \\ 
 293 & 20130531 1952 & M1.0 & N12E42 &  1060 &    336 &    5.9E+05 & [10-30] &  6.7 &    4.5E+30 &     1.06 &     2.97 \\ 
 296 & 20130621 0230 & M2.9 & S14E73 &  5068 &    912 &    3.7E+06 & [10-25] &  7.4 &    1.2E+32 &     0.12 & ... \\ 
 297 & 20130623 2048 & M2.9 & S18E63 &  1132 &   2160 &    2.7E+06 & [10-30] &  5.1 &    3.0E+31 &    0.028 & ... \\ 
 298 & 20130703 0700 & M1.5 & S14E82 &  1548 &   1008 &    1.9E+06 & [10-30] &  5.1 &    2.1E+31 &     0.26 & ... \\ 
 299 & 20130812 1021 & M1.5 & S21E17 &  1536 &    976 &    2.2E+06 & [10-25] &  6.5 &    8.5E+31 &    0.071 &     5.07 \\ 
 303 & 20131011 0701 & M1.5 & N21E87 &  1124 &    688 &    8.2E+05 & [10-30] &  4.7 &    5.1E+30 &     0.14 & ... \\ 
 304 & 20131013 0012 & M1.7 & S22E17 &  1416 &    400 &    8.7E+05 & [ 8-25] &  5.5 &    2.0E+31 &     0.35 &     0.25 \\ 
 306 & 20131015 2331 & M1.3 & S21W22 &  1720 &    912 &    1.0E+06 & [10-30] &  6.8 &    2.0E+31 &     0.19 &     0.52 \\ 
 307 & 20131017 1509 & M1.2 & S09W63 &  1696 &    352 &    2.1E+06 & [10-30] &  5.7 &    5.1E+30 &     1.11 & ... \\ 
 308 & 20131022 0014 & M1.0 & N08E20 &  1068 &    752 &    1.1E+06 & [10-30] &  4.8 &    5.1E+31 &    0.073 &     0.34 \\ 
 311 & 20131023 2041 & M2.7 & N08W06 &  3368 &   1904 &    5.4E+06 & [10-30] &  3.2 &    1.9E+31 &     0.33 &     0.11 \\ 
 312 & 20131023 2333 & M1.4 & N09W08 &  2000 &   1136 &    1.4E+06 & [10-35] &  5.2 &    1.5E+30 &     2.84 &    0.004 \\ 
 313 & 20131023 2358 & M3.1 & N09W09 &  1452 &   2416 &    6.6E+06 & [10-25] &  7.9 &    3.7E+31 &     0.20 &     0.25 \\ 
 317 & 20131025 0248 & M2.9 & S07E76 &  3164 &   1840 &    5.5E+06 & [10-30] &  5.8 &    7.0E+31 &     0.16 & ... \\ 
 318 & 20131025 0753 & X1.7 & S08E73 &   676 &  10409 &    1.1E+07 & [10-25] &  9.0 &    3.4E+33 &   0.0032 & ... \\ 
 320 & 20131025 1451 & X2.1 & S06E69 &  3568 &  16678 &    6.5E+07 & [10-25] & 10.8 &    3.4E+32 &    0.072 & ... \\ 
 321 & 20131025 1702 & M1.3 & S08E67 &  2052 &    847 &    3.2E+06 & [10-30] &  4.2 &    8.8E+30 &     0.28 & ... \\ 
 324 & 20131026 0559 & M2.3 & S08E59 &  1880 &   2032 &    3.4E+06 & [10-20] &  4.9 &    1.4E+31 &     0.24 & ... \\ 
 325 & 20131026 0917 & M1.5 & S08E59 &  1060 &    320 &    6.5E+05 & [10-30] &  5.4 &    4.3E+30 &     0.67 & ... \\ 
 326 & 20131026 1048 & M1.8 & S06E59 &  1176 &    320 &    1.0E+06 & [10-30] &  7.3 &    5.5E+31 &     0.14 & ... \\ 
 328 & 20131026 1949 & M1.0 & S08E51 &  1940 &    272 &    6.2E+05 & [10-25] &  6.6 &    1.4E+30 &     0.60 & ... \\ 
 330 & 20131028 0141 & X1.0 & N05W72 &  2376 &   9863 &    3.1E+07 & [10-20] &  6.9 &    1.9E+32 &     0.12 & ... \\ 
 332 & 20131028 1132 & M1.4 & S14W46 &  3956 &    309 &    2.3E+06 & [10-30] &  8.6 &    5.0E+31 &     0.11 & ... \\ 
 334 & 20131028 1446 & M2.7 & S08E27 &  2600 &   2288 &    8.8E+06 & [10-30] &  6.5 &    1.3E+32 &     0.24 &     2.53 \\ 
 336 & 20131028 2048 & M1.5 & N07W83 &  1748 &   1200 &    1.5E+06 & [10-30] &  7.0 &    4.9E+31 &    0.037 & ... \\ 
 340 & 20131102 2213 & M1.6 & S12W12 &   768 &   1200 &    1.6E+06 & [10-30] &  6.4 &    6.4E+31 &    0.037 &     0.73 \\ 
 343 & 20131105 1808 & M1.0 & S12E47 &  1124 &    688 &    8.2E+05 & [10-30] &  6.1 &    1.2E+31 &     0.12 & ... \\ 
 345 & 20131106 1339 & M3.8 & S09E35 &  1936 &   2928 &    6.0E+06 & [10-30] &  6.0 &    1.1E+32 &    0.043 &     0.64 \\ 
 347 & 20131107 0334 & M2.3 & S08E26 &  1436 &   1776 &    1.7E+06 & [10-25] &  5.1 &    4.8E+31 &     0.15 &     0.13 \\ 
 351 & 20131110 0508 & X1.1 & S11W17 &  3284 &   9303 &    1.3E+07 & [10-30] &  8.0 &    1.3E+33 &    0.043 &     4.95 \\ 
 352 & 20131111 1101 & M2.4 & S17E74 &  3068 &   1264 &    6.6E+06 & [10-30] &  7.3 &    2.3E+32 &    0.032 & ... \\ 
 353 & 20131113 1457 & M1.4 & S20E46 &  1400 &    592 &    1.3E+06 & [10-30] &  8.1 &    4.1E+31 &     0.25 & ... \\ 
 354 & 20131115 0220 & M1.0 & N07E53 &  1252 &    656 &    9.0E+05 & [10-30] &  6.9 &    4.5E+31 &    0.086 & ... \\ 
 357 & 20131117 0506 & M1.0 & S19W41 &  1208 &    592 &    5.4E+05 & [10-25] &  7.0 &    2.4E+31 &    0.025 &     0.20 \\ 
 359 & 20131121 1052 & M1.2 & S14W89 &  1248 &    448 &    2.1E+06 & [10-25] &  5.7 &    3.5E+31 &    0.040 & ... \\ 
 360 & 20131123 0220 & M1.1 & N13W58 &  2584 &    432 &    1.6E+06 & [10-30] &  4.1 &    8.2E+31 &    0.034 & ... \\ 
 363 & 20131219 2306 & M3.5 & S16E89 &  2304 &   2160 &    5.5E+06 & [10-30] &  7.0 &    2.8E+32 &    0.055 & ... \\ 
 364 & 20131220 1135 & M1.6 & S16E78 &  4272 &    336 &    2.1E+06 & [10-30] &  5.3 &    9.6E+30 &     0.37 & ... \\ 
 365 & 20131222 0805 & M1.9 & S17W51 &  1788 &   1776 &    2.2E+06 & [10-30] &  8.2 &    4.2E+31 &    0.054 & ... \\ 
 366 & 20131222 0833 & M1.1 & S17W52 &  1956 &    320 &    5.9E+05 & [10-25] &  5.8 &    8.4E+30 &     0.32 & ... \\ 
 367 & 20131222 1424 & M1.6 & S16E44 &  2532 &    416 &    1.7E+06 & [10-30] &  5.3 &    5.1E+31 &     0.14 &     0.79 \\ 
 368 & 20131222 1506 & M3.3 & S17W55 &  1328 &   1968 &    3.4E+06 & [10-30] &  6.4 &    3.0E+31 &     0.37 & ... \\ 
 377 & 20140103 1241 & M1.0 & S04E52 &  1000 &     30 &    9.5E+04 & [10-30] &  3.6 &    2.0E+29 &     5.44 & ... \\ 
 379 & 20140104 1016 & M1.3 & S05E49 &  2888 &    400 &    2.0E+06 & [10-30] &  5.4 &    2.0E+31 &     0.23 & ... \\ 
 382 & 20140107 0349 & M1.0 & N07E07 &  1432 &    880 &    8.1E+05 & [10-25] &  7.3 &    2.9E+31 &    0.051 &     0.39 \\ 
 383 & 20140107 1007 & M7.2 & S13E13 &  2000 &   7967 &    2.8E+07 & [16-30] &  8.7 &    2.4E+33 &   0.0076 &     4.47 \\ 
 385 & 20140108 0339 & M3.6 & N11W88 &  2016 &   2672 &    4.4E+06 & [10-30] &  5.4 &    1.5E+32 &   0.0057 & ... \\ 
 386 & 20140113 2148 & M1.3 & S08W75 &   660 &   1456 &    7.8E+05 & [10-25] &  8.3 &    3.4E+31 &    0.013 & ... \\ 
 387 & 20140127 0105 & M1.0 & S16E88 &  2860 &    272 &    1.5E+06 & [10-30] &  3.3 &    4.5E+29 &    10.74 & ... \\ 
 389 & 20140127 2205 & M4.9 & S14E88 &  1880 &   4129 &    5.8E+06 & [10-30] &  9.8 &    2.6E+32 &   0.0064 & ... \\ 
 393 & 20140128 1233 & M1.3 & S15E79 &  1708 &    560 &    1.2E+06 & [10-30] &  9.1 &    1.7E+31 &    0.020 & ... \\ 
 395 & 20140128 2204 & M2.6 & S14E74 &  1112 &   1968 &    2.5E+06 & [10-30] &  6.5 &    1.3E+32 &   0.0055 & ... \\ 
\enddata
\end{deluxetable}
\clearpage


\begin{figure}
\plotone{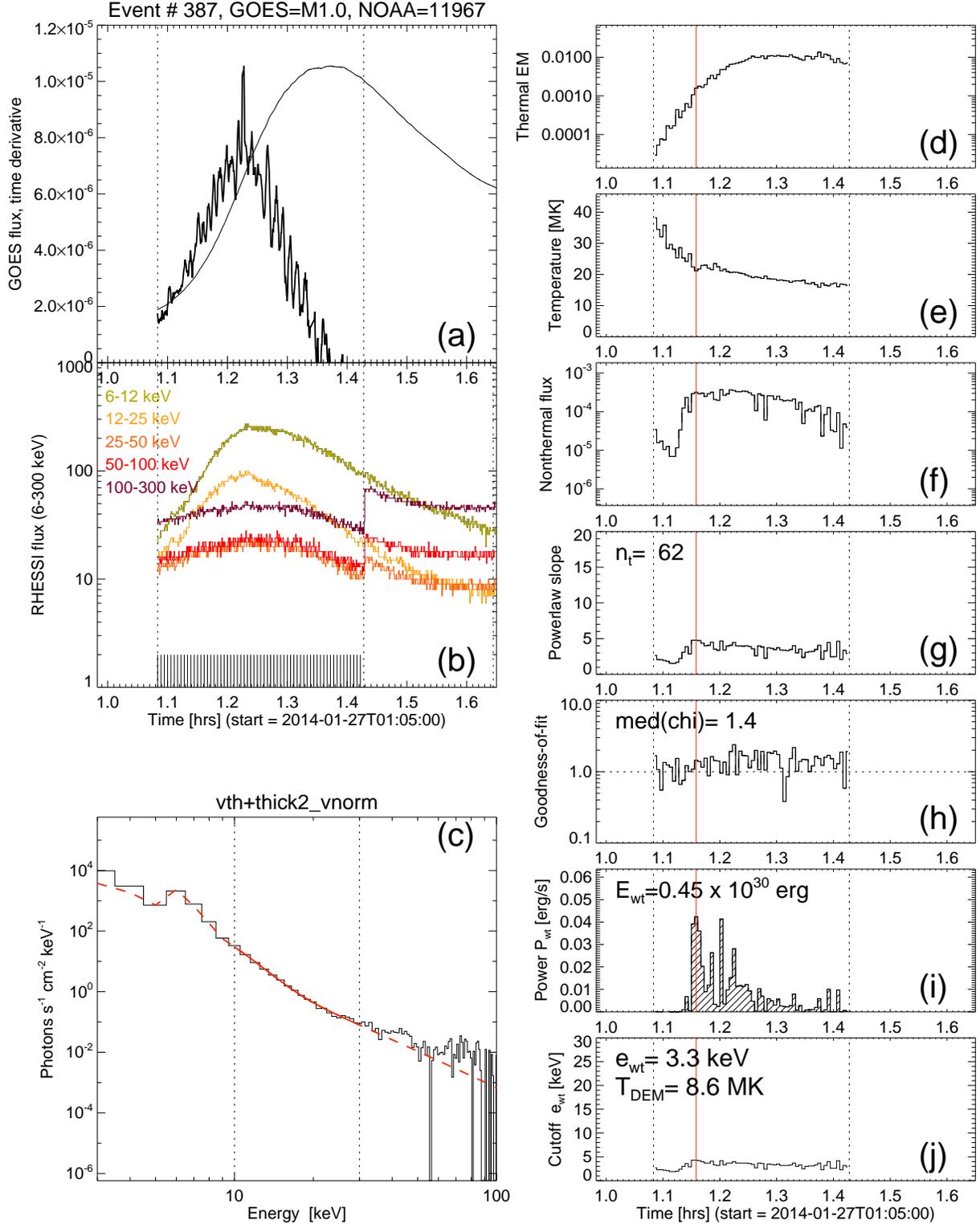}
\caption{Data analysis of the small flare event \#387, GOES M1.0-class,
observed on 2014-Jan-27, 01:05 UT: 
(a) the GOES 1-8 \ang\ flux and time derivative; 
(b) RHESSI quick-look time profiles in 5 energy channels in the range of 6-300 keV; 
(c) the spectral fit at the peak time of the nonthermal power
$e_{\mathrm{wt}}(t)$ (red),
(d) the thermal emission measure $EM(t)$;
(e) the temperature evolution $T_e(t)$;
(f) the nonthermal photon flux $I_{\mathrm{nt}}(t)$;
(g) the power law slope $\delta(t)$;
(h) the goodness-of-fit $\chi(t)$; 
(i) the nonthermal power $P_{\mathrm{wt}}(t)$;
(j) the low-energy cutoff $e_{\mathrm{wt}}(t)$.
The dotted lines indicate fitting ranges and the vertical red lines
indicate the peak time of the nonthermal power.}
\end{figure}
\clearpage

\begin{figure}
\plotone{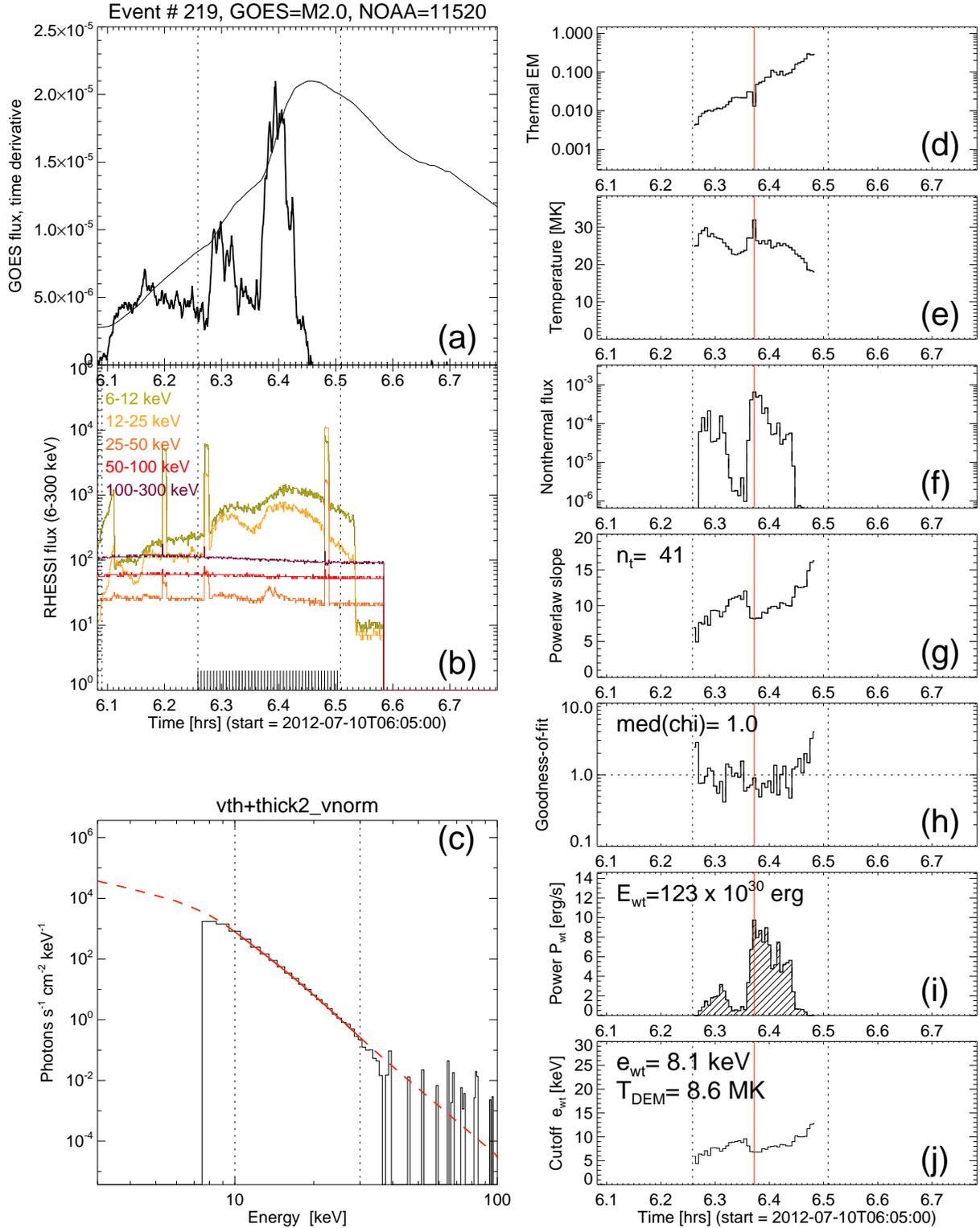}
\caption{The data analysis of a medium-size flare event \#219, GOES M2.0-class,
observed on 2012-Jul-10, 06:05 UT, otherwise similar representation as in Fig.~1.}
\end{figure}
\clearpage

\begin{figure}
\plotone{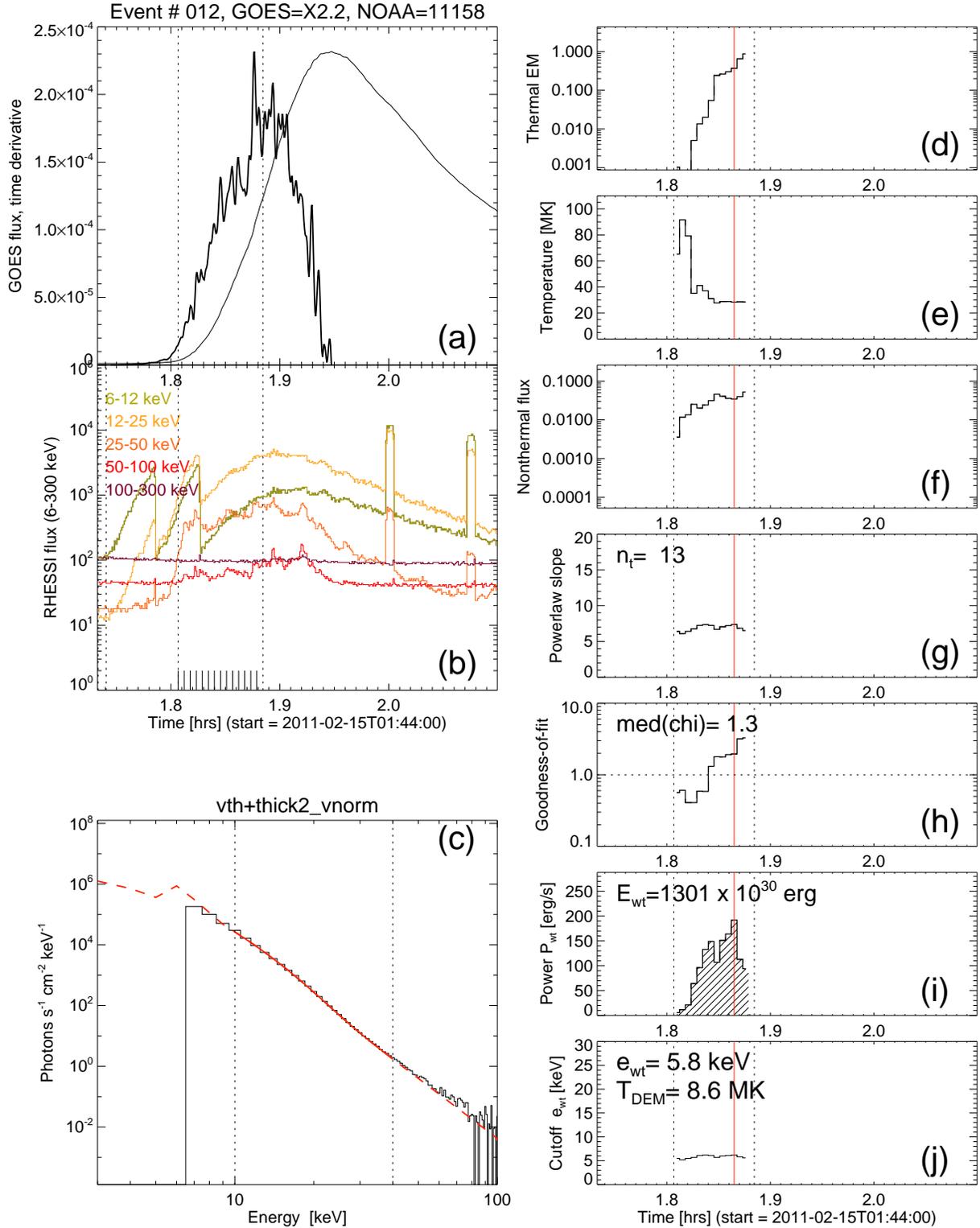}
\caption{The data analysis of a large flare event \#12, GOES X2.2-class,
observed on 2011-Feb-15, 01:44 UT, otherwise similar representation 
as in Figs.~1 and 2.}
\end{figure}
\clearpage

\begin{figure}
\plotone{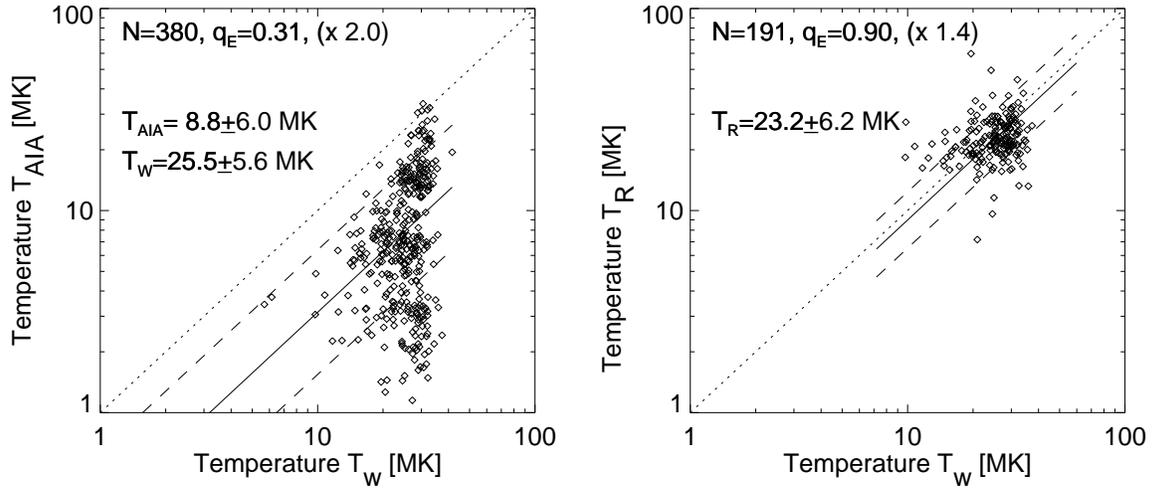}
\caption{Comparison of three different temperature definitions: 
the DEM peak temperature $T_{\mathrm{AIA}}$ as a function of the 
DEM-weighted temperature $T_w$ as measured in Paper II (left-hand 
panel), and the time-averaged RHESSI temperature $T_{R}$ as a function 
of $T_w$ (right-hand panel). The (logarithmically) averaged temperature 
ratio is indicated with a solid line, the logarithmic standard deviation 
with two dashed lines, and the unity ratio with a dotted line.}
\end{figure}
\clearpage

\begin{figure}
\plotone{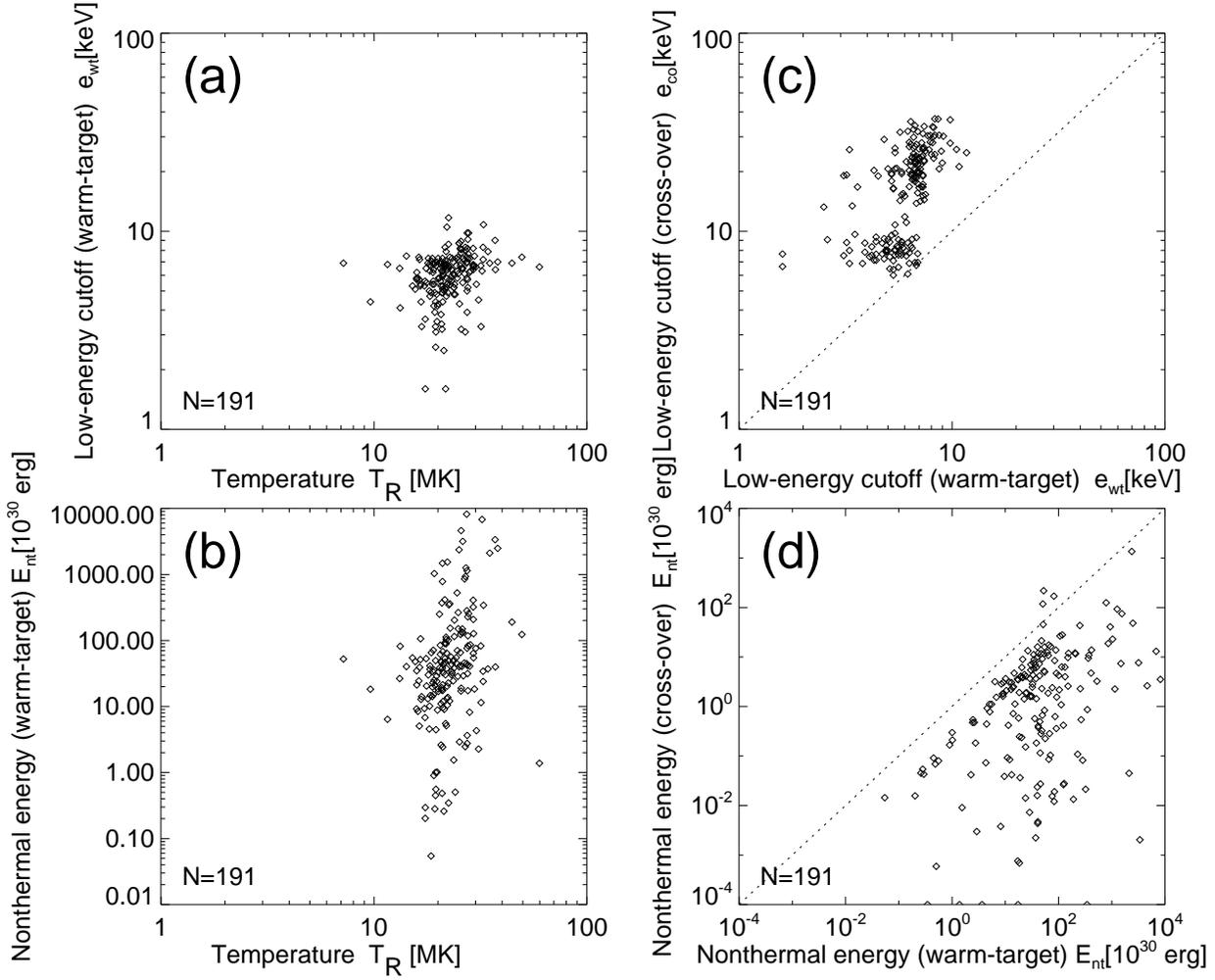}
\caption{Functional dependence of the low-energy cutoff $e_{\mathrm{wt}}$ (a),
and the (warm-target) nonthermal energy $E_{\mathrm{nt}}$ (b) as a 
function of the RHESSI temperature $T_{R}$. Scatterplots of
the low-energy cutoffs (c) and the nonthermal energies (d) 
are shown between the warm-target and the cross-over model.
The diagonal dotted lines (in right-hand panels) indicate equivalence.
Note that the cross-over method yield systematically larger cutuoff 
energies and smaller nonthermal energies than the warm-target model.}
\end{figure}
\clearpage

\begin{figure}
\plotone{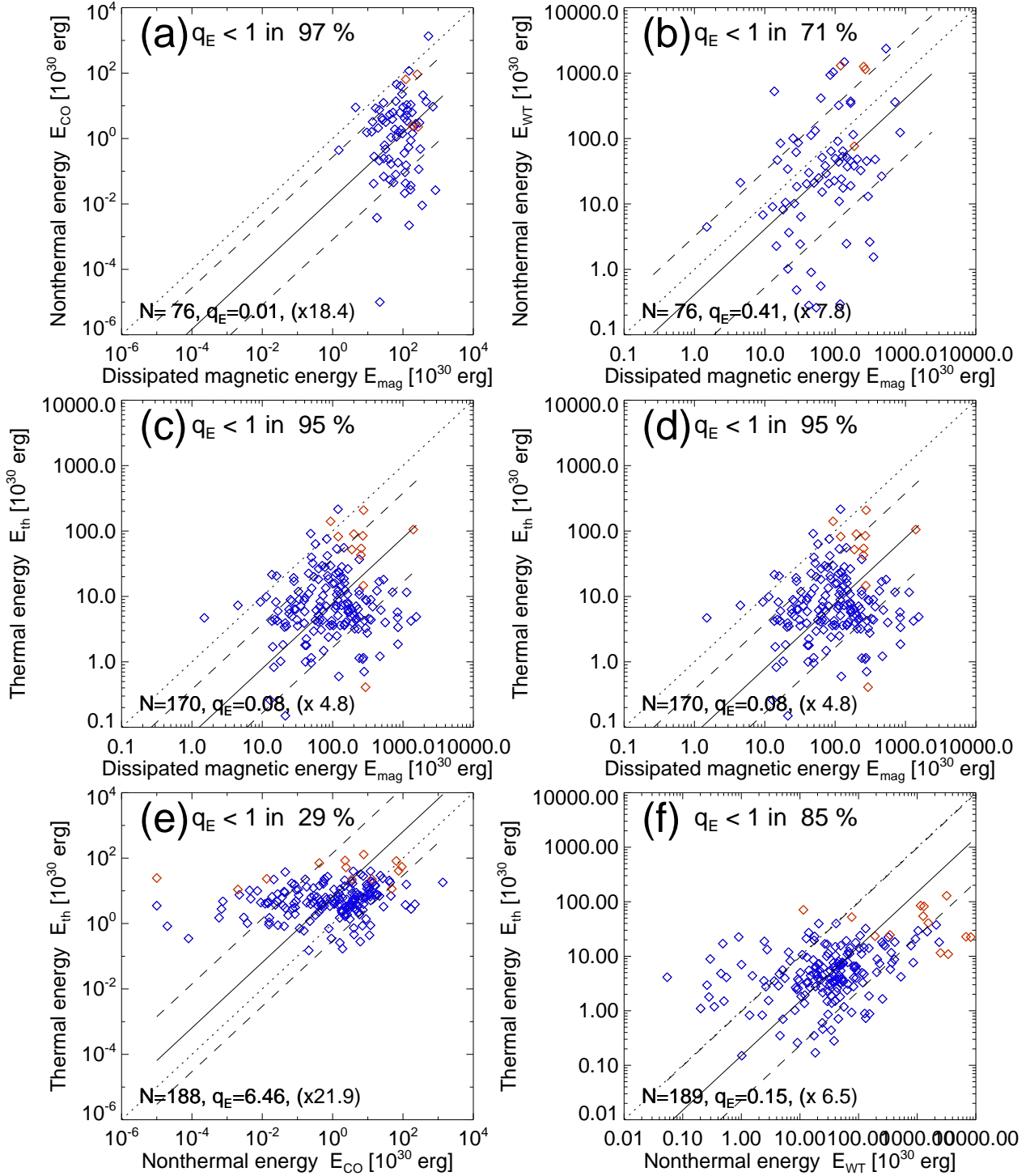}
\caption{Scatter plots are shown between the dissipated magnetic
energies $E_{mag}$ (calculated in Paper 1), the thermal energies
$E_{th}$ (calculated in Paper 2), and the non-thermal energies 
calculated here, using the cross-over method (left panels a,e) 
and the warm-target model (right panels b,f). 
The mean ratios (by averaging the logarithmic values)
(solid lines) are indicated in each panel with the standard 
deviations (two dashed lines and multiplier marked with $\times$), 
and the unity ratio (dotted line). The color code indicates
X-class (red) and M-class flares (blue).}
\end{figure}
\clearpage

\begin{figure}
\plotone{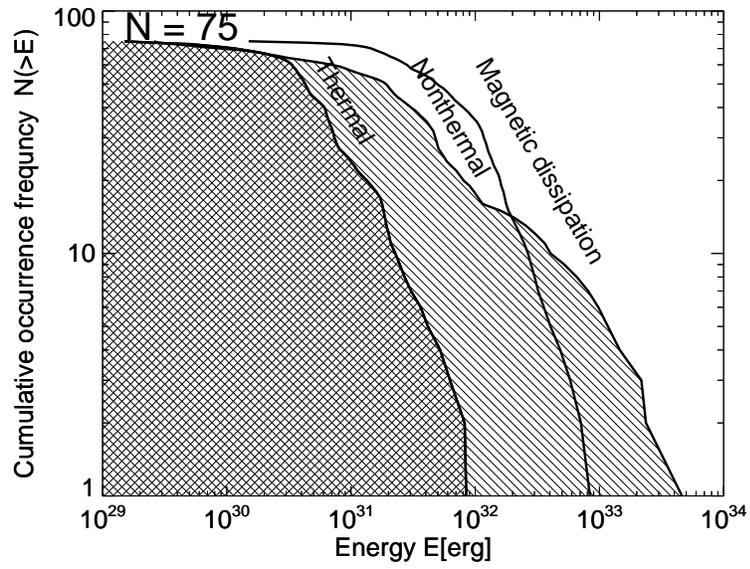}
\caption{Cumulative occurrence frequency distributions of thermal,
nonthermal, and dissipated magnetic energies in 75 M and
X-class flare events simultaneously observed with HMI, AIA, 
and RHESSI.}
\end{figure} 
\clearpage

\begin{figure}
\plotone{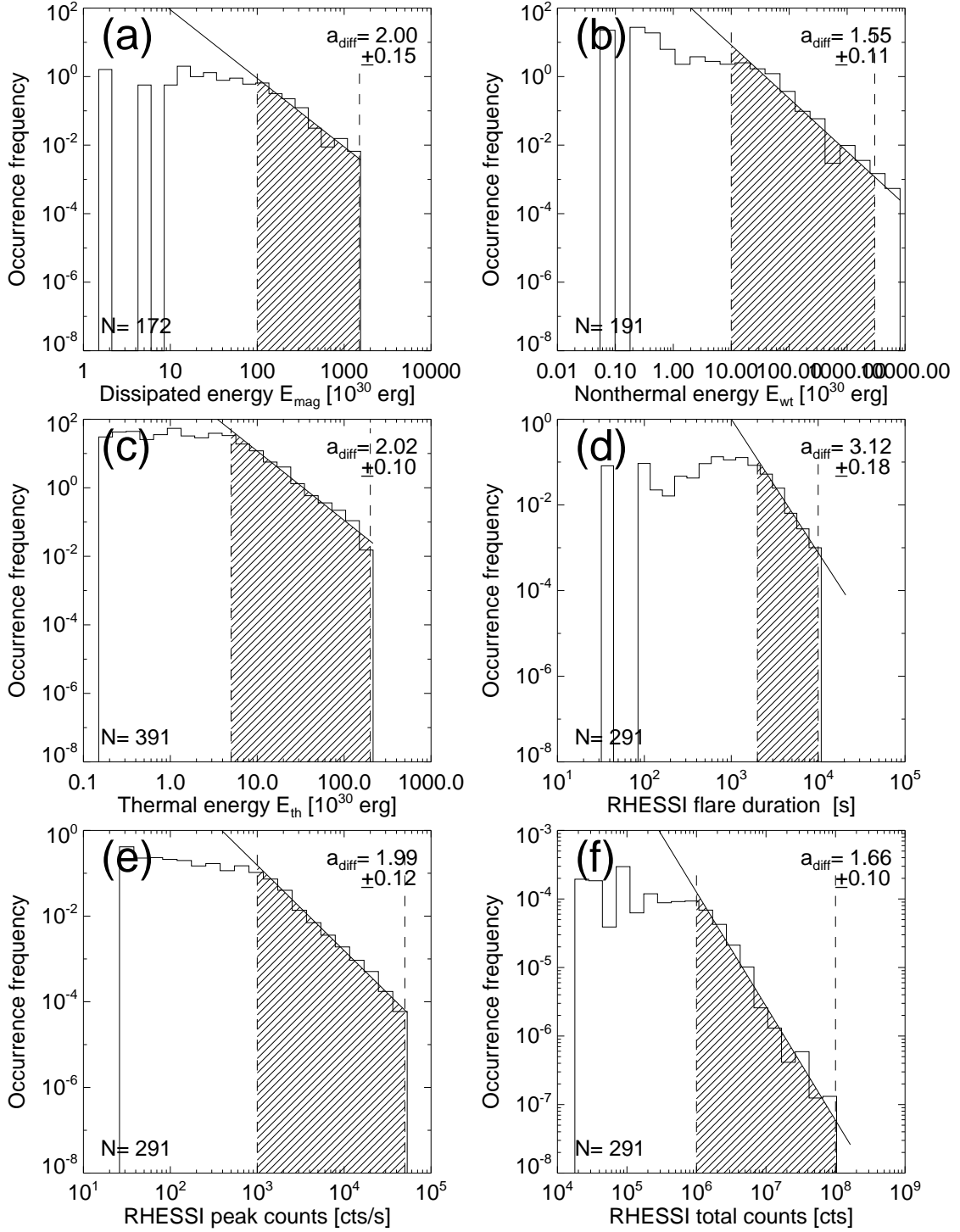}
\caption{Occurrence frequency distributions of dissipated magnetic 
energies (a), RHESSI nonthermal energies (b), AIA thermal energies 
(c), RHESSI flare durations (d), RHESSI peak counts (e), and RHESSI 
total counts (f). Only the histogram parts with complete sampling 
(hatched areas) are fitted with a power law function.}
\end{figure} 
\clearpage

\begin{figure}
\plotone{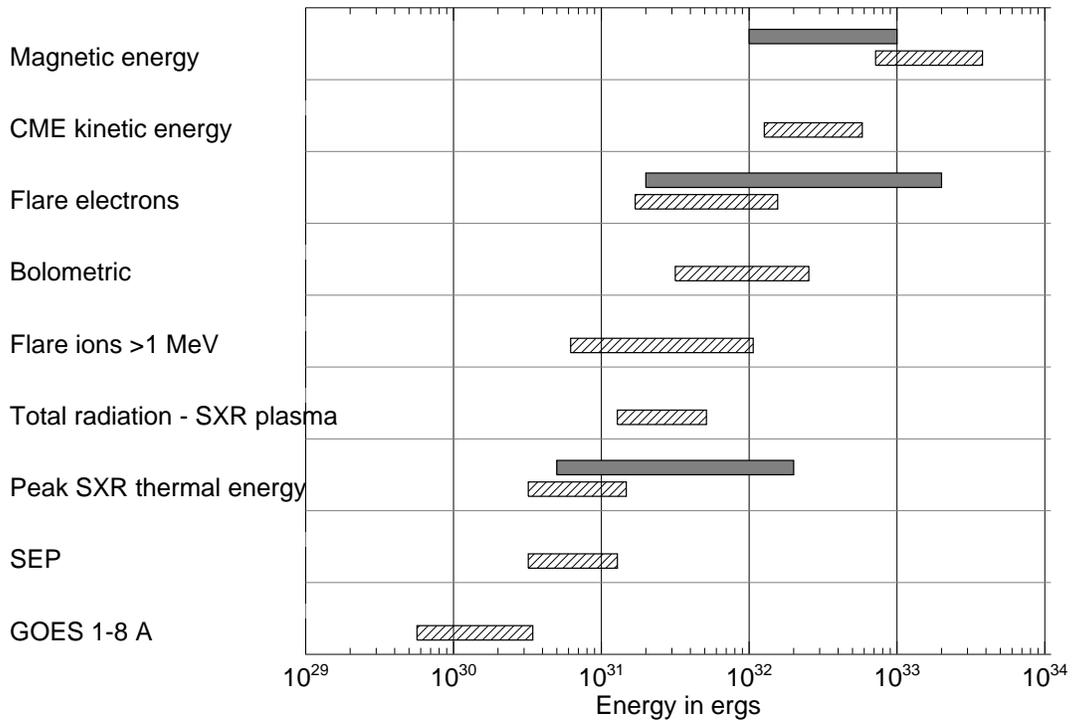}
\caption{Bar chart showing the logarithmic ranges of energy distributions
according to the study of 38 events in Emslie et al.~(2012) (hatched boxes).
For comparison, 
the magnetic energies in 172 events (Aschwanden et al.~2014), the thermal
energies in 391 events (Aschwanden et al.~2015), and the nonthermal
energies in this study here are shown (all in grey boxes). 
The grey boxes exclude incompletely-sampled ranges.}
\end{figure}
\clearpage

\end{document}